\documentclass[apj]{emulateapj}

\shorttitle{H$\alpha$ Morphologies of Virgo Cluster Spirals}

\journalinfo{The Astrophysical Journal}
\submitted{Accepted 2004 June 3}

\received{2002 August 23}
\begin{document}

\title{H$\alpha$ Morphologies and Environmental Effects in
Virgo Cluster Spiral Galaxies}

\author{Rebecca A. Koopmann}
\affil{Union College}
\affil{Department of Physics, Schenectady, NY 12308}
\email{koopmanr@union.edu}

\author{Jeffrey D. P. Kenney}
\affil{Astronomy Department}
\affil{Yale University, P.O. Box 208101, New Haven, CT 06520-8101}
\email{kenney@astro.yale.edu}

\begin{abstract}

We describe the various H$\alpha$ morphologies of Virgo Cluster and isolated 
spiral galaxies, 
and associate the  H$\alpha$ morphologies with
the types of environmental interactions which have altered 
the cluster galaxies.
The spatial distributions of  H$\alpha$ and R-band emission
are used to divide
the star formation morphologies of the 52 Virgo Cluster spirals into several 
categories: normal (37\%), anemic (6\%),  enhanced (6\%), and 
(spatially) truncated (52\%). 
Truncated galaxies are further subdivided based on their 
inner star formation rates into truncated/normal (37\%), 
truncated/compact (6\%),
truncated/anemic (8\%), and truncated/enhanced (2\%).
The fraction of anemic galaxies is relatively small (6-13\%) in both 
environments, 
suggesting that starvation is not a major factor in the 
reduced star formation rates of Virgo spirals.
The majority of Virgo spiral galaxies have their H$\alpha$ disks 
truncated (52\%),
whereas truncated H$\alpha$ disks are rarer in isolated galaxies (12\%).
Most of the H$\alpha$-truncated galaxies have 
relatively undisturbed stellar disks 
and normal-to-slightly enhanced inner disk star formation rates, suggesting
that ICM-ISM stripping is the
main mechanism causing the reduced star formation rates of Virgo spirals.
Several of the truncated galaxies are peculiar, with enhanced central star
formation rates, disturbed stellar disks,
and bar-like distributions of luminous HII complexes inside the central 1 kpc,
but no star formation beyond, suggesting
recent tidal interactions or minor mergers have also influenced their
morphology. 
Two highly-inclined H$\alpha$-truncated spirals have numerous
extraplanar HII regions, and are likely
in an active phase of ICM-ISM stripping.
Several spirals have one-sided H$\alpha$ enhancements
at the outer edge of their truncated H$\alpha$ disks, suggesting 
modest and local enhancements in their star formation rates 
due to ICM-ISM interactions.
Low-velocity tidal interactions and perhaps 
outer cluster HI accretion seem to be
the triggers for enhanced global star formation in four Virgo galaxies.
These results indicate that 
most Virgo spiral galaxies experience ICM-ISM stripping,
many experience significant tidal effects, 
and many experience both.

\end{abstract}


\section{Introduction}

Many kinds of environmental interactions have been proposed to affect
galaxies in clusters, yet it is still unknown
which processes actually occur and the key details
of what happens in the different kinds of interactions.

Processes which tend to affect the gas content of the galaxy, but not
the existing stellar content, include
(i) intracluster medium - interstellar medium (ICM-ISM) interactions 
(Gunn \& Gott 1972; Nulsen 1982;
Schulz \& Struck 2001; Vollmer et al. 2001; van Gorkom 2004)
which selectively remove ISM gas from galaxies
and might also compress the ISM and trigger star formation,
(ii) gas accretion, which would presumably occur in the cluster outskirts, and
(iii) starvation or strangulation, the stripping of
gas from a galaxy's surroundings which might otherwise have accreted onto
the galaxy (Larson, Tinsley, \& Caldwell 1980;
Balogh, Navarro, \& Morris et al. 2000).
Tidal or gravitational effects (Toomre \& Toomre 1972; reviews by Struck
1999 and Mihos 2004),
which affect both stars and gas, are
also likely to be important, and
include
(i) low-velocity tidal interactions and mergers, which can occur
both inside and outside of clusters,
(ii) high-velocity tidal interactions and collisions, which
occur almost exclusively in clusters (e.g., Moore, Lake, \& Katz 1998), and
(iii) tidal interactions between galaxies and the cluster as a whole,
or sub-units within clusters (Byrd \& Valtonen 1990).

There have been a number of recent studies of environmental processes in
nearby and distant
cluster galaxies with authors reaching a variety of conclusions
(see review by O'Connell 1999)
concerning star formation histories and lenticular formation.
They can be roughly divided into four categories: 
(i) those which suggest that the major influence
on spirals in clusters is ICM-ISM stripping, which reduces star
formation mainly by simple truncation, although modest bursts
may happen in at least some galaxies (e.g., Abraham et al. 1996;
Bravo-Alfaro et al. 2000, 2001; Jones, Smail, \& Couch 2000; 
Couch et al. 1998).
(ii) those which suggest that
starbursts caused by tidal interactions with the cluster or other
galaxies have had a major effect on the evolution of a significant
number of galaxies (e.g., Moss \& Whittle 2000; Caldwell et al. 1996;
Caldwell, Rose, \& Dendy 1999;
Henrikson \& Byrd 1996; van Dokkum et al. 1999; Rose et al. 2001).
(iii)  those which suggest that the reduction in cluster galaxy star formation
has been gradual (Balogh et al. 1999) or occurs also in the outer cluster
(Treu et al. 2003), more consistent with starvation than (complete) 
ICM-ISM stripping.
(iv) those which suggest that at least 2 of the above processes, e.g.,
ICM-ISM stripping and tidal interactions are necessary to explain
observed star formation properties (Poggianti et al. 1999; 2001).
It is interesting that different
studies disagree about the main environmental effects even when
analyzing the same cluster!
Taken together, the results of these
studies indicate that the properties of cluster galaxies may be
determined
by a variety of environmental interactions over a Hubble time (e.g., Miller
1988; Oemler 1992; Moore et al. 1998) and may vary between
clusters, and over time.
On the other hand,
the complexity and disagreement shows that detailed studies 
of star formation in cluster galaxies are needed.

In this paper we 
try to establish a link between the H$\alpha$ morphologies of Virgo Cluster 
spirals
and the types of interaction(s) which may have affected them.
The spatial distribution of  H$\alpha$ emission is
a sensitive indicator of some types of environmental interactions,
since it traces part of the ISM, as well as star formation activity.
We have chosen to study galaxies
in the Virgo Cluster, the nearest moderately rich cluster, so that
we can study cluster interactions at the highest possible resolution.

This paper is one of a series based on our
H$\alpha$ and R imaging survey of Virgo Cluster and isolated spirals
galaxies.
We present the observational
data for the Virgo galaxies in Koopmann et al. (2001, hereafter PI)
and that for the isolated galaxies in Koopmann \& Kenney (2004, in prep,
hereafter PII).
These papers include H$\alpha$ and R images and radial profiles
for all galaxies.
In a companion paper (Koopmann \& Kenney 2004, hereafter PIII),
we present a detailed comparison of H$\alpha$ radial distributions
and star formation rates of the Virgo and isolated galaxies,
finding that truncation of the star-forming disk is the main reason for
reduced Virgo star formation rates, but that the star formation rates of
individual Virgo spirals range from reduced to enhanced.
In this paper, we describe more fully the different types of
H$\alpha$ radial distributions and morphologies,
and use this and other information
in an attempt to determine the types of environmental effects which have
altered the Virgo spirals.

\section{Star formation morphologies in the Virgo Cluster}
\label{sfmorph}

In Papers I, II, and III, we present R and H$\alpha$ images, radial
profiles,
integrated fluxes, and concentration indices for 55 Virgo Cluster and
29 isolated S0-spiral galaxies which are
brighter than $M_{\rm B}$=-18, and have
inclinations less than 75\arcdeg.
In this paper we exclude the S0's, and consider a sample
of 52 Virgo Cluster and 24 isolated  Sa-Scd galaxies.
In Paper III, galaxies were compared via the Hubble types, as well as
the several quantitative measures of the
radial distributions and relative amounts of R and H$\alpha$,
which we also use in this paper. The central R light concentration,
$C30$, is the inverse ratio of R-band flux within the 24
magnitudes per arcsec$^{-2}$ isophote, $r_{24}$,
and the 0.3 $r_{24}$ isophote. This is a measure of the stellar
light central concentration, and a tracer of the bulge-to-disk ratio.
CH$\alpha$ is an analogous quantity for the H$\alpha$ concentration,
i.e. the ratio of H$\alpha$ flux within 0.3$r_{24}$ to that within
$r_{24}$.
The normalized massive star formation rate (NMSFR) is 
the ratio of H$\alpha$ flux to R flux, measured within the same aperture,
and is similar to an equivalent width.

In Paper III, we show that truncation of the star forming disk
is the most common cause of reduction in star formation in the
Virgo Cluster by examining radial profiles and integrated NMSFRs. In
this paper, we wish to quantify the numbers of galaxies affected by 
truncation, as well as other types of star formation reduction or enhancement, 
using objective, 
comparative measures of a galaxy's H$\alpha$ and R radial profile.
We base our definitions on the values of NMSFR for the 5 radial bins
($r < $0.1$r_{24}$, 0.1$r_{24} < r  < $0.3$r_{24}$, 
0.3$r_{24} < r  < $0.5$r_{24}$, 0.5$r_{24} < r  < $0.7$r_{24}$,
and 0.7$r_{24} < r < $1.0$r_{24}$) shown in Figure 8 of Paper III. 
In order to define a `normal' NMSFR for isolated galaxies,
we binned the isolated sample spiral galaxies (Sa-Scd) into low concentration 
($C30 \leq$ 0.40) and high concentration ($C30 >$ 0.40) classes. Each
bin contained 12 galaxies. We computed the low and high concentration median 
isolated NMSFR for each of the radial bins and compared these values to
the NMSFRs of individual galaxies of high and low concentrations in each
radial bin. We then looked for trends of enhancement or reduction over
multiple bins of the star-forming disk of each galaxy in order to define
several star formation morphology classes. A second iteration was made removing
the four isolated galaxies which fit into categories other than `normal'.
Table~\ref{medsfrs} presents the isolated median rates (in units of equivalent
width) derived for each bin. 

\begin{deluxetable}{rcc}
\tablecaption{Median Isolated Spiral H$\alpha$ Equivalent Widths}
\tablewidth{0pt}
\tablehead{
\colhead{Radial Bin}&
\colhead{EW (\AA)}&
\colhead{EW (\AA)}\\
\colhead{}&
\colhead{0.20 $\le C30 \le$ 0.40}&
\colhead{0.41 $\le C30 \le$ 0.60}}
\startdata
0 $<$ r $<$ 0.1$r_{24}$ & 21 & 6 \\
0.1$r_{24} <$ r $<$ 0.3$r_{24}$ & 19 & 10 \\
0.3$r_{24} <$ r $<$ 0.5$r_{24}$ & 32 & 19 \\
0.5$r_{24} <$ r $<$ 0.7$r_{24}$ & 34 & 15 \\
0.7r $_{24} <$ r $<$ 1.0$r_{24}$ & 30 & 13 \\
Total & 35 & 15\\
\enddata
\label{medsfrs}
\end{deluxetable}

We define the following star formation morphology classes:

(N) \bf Normal: \rm spirals with global NMSFRs within a factor of 3 of
the isolated global medians, as well as NMSFRs within a factor of 5
of the isolated median in the innermost (r$<$0.1$r_{24}$) bin, and NMSFRs
within a factor of 3 in at least three of the other four radial bins. 
In most cases, these galaxies are within a factor of 3 of the isolated
median for each radial bin. However, enhancements/reductions of a factor
greater than 3 do occur within single bins for several galaxies, due to
the `bumpiness' of the H$\alpha$ distribution, which can
be caused by barred regions, rings of star formation, and circumnuclear
activity. In two cases (1 isolated, 1 Virgo) 
the innermost (circumnuclear) bin rate is enhanced or reduced by 3-5 times.
We have elected to include these galaxies in the normal category because
the circumnuclear bin contains the center of the galaxy and is therefore
most susceptible to color-dependent continuum subtraction errors and 
contamination by the [NII] line. In a few cases (6 isolated, 1 Virgo),
one of the other four bins is enhanced or reduced by slightly more than a 
factor of 3. Examination of the H$\alpha$ image shows that these 
reductions/enhancements are due either to bars or rings of star formation. 

(A) \bf Anemic: \rm spirals which have low but measurable NMSFR across
all radial bins.  The requirements to fit into this class are a
global NMSFR reduced by at least a factor of 3 together with 
a detectable NMSFR in all bins which is reduced by at least a
factor of 2. The detectability of the NMSFR was 
confirmed by inspection of the radial H$\alpha$ profile and the
H$\alpha$ images. 

(E) \bf Enhanced: \rm spirals with global NMSFRs enhanced by at least
a factor of three compared to isolated medians. 

(T) \bf Truncated: \rm  spirals with a sharp cutoff in the star-forming
disk, with a reduction in NMSFR 
of at least 10 times in the outermost (0.7 - 1.0$r_{24}$) bin. 
This requirement selects galaxies with truncation radii smaller than
about 0.8 $r_{24}$.  

About half (52\%) 
of the Virgo Cluster spirals have truncated star-forming disks. We 
subdivide these galaxies based on their inner disk star formation rates and
morphologies into the following classes:

(T/N) \bf Truncated/Normal: \rm 
spirals with `normal' inner disk NMSFRs interior to the 
truncation radius, where `normal' is defined as above.
Interior to the truncation radius, the
H$\alpha$ surface brightness distribution is
similar to that of the large scale disk in the R band.
(This is in contrast to the Truncated/Compact galaxies,
described in the next class.)

Within this class, one can further distinguish between 
those with severe truncation (truncation radius less than 0.4$r_{24}$),
or moderate truncation (truncation radius between 0.4-0.8$r_{24}$).
Galaxies with mild truncation (truncation radius greater than 0.8$r_{24}$) 
also exist,
but these are hard to unambiguously identify with our data, so in this paper 
are in the Normal class.

(T/C) \bf Truncated/Compact: \rm spirals with 
circumnuclear (r$ < 0.1r_{24}$ bin) 
NMSFRs enhanced by at least a factor of 5, together with
a reduction of at least a factor of 10 in the outermost 2 bins.
In these galaxies, the H$\alpha$ radial surface brightness profile 
is much steeper than that of the large scale disk at every radius,
and closer to that of a "bulge",
with much of the emission concentrated in the central 1 kpc.
This class of H$\alpha$ morphology is distinct from
a Truncated/Normal galaxy with a severe truncation, which has an
H$\alpha$ radial profile similar to that of the large scale disk 
inward of the truncation radius. 
In Truncated/Compact galaxies,
most of the H$\alpha$ emission arises from a
non-axisymmetric distribution of luminous circumnuclear HII complexes. 

\noindent

(T/A) \bf Truncated/Anemic: \rm spirals with
detectable NMSFRs in the innermost bins which are reduced 
in level by at least a factor of 2 compared to the innermost isolated
bin medians. 

(T/E) \bf Truncated/Enhanced: \rm spirals with NMSFRs enhanced by
at least a factor of 3 above isolated medians in at least two
bins interior to the truncation radius. 

Table~\ref{tabclass} gives the percentages of Virgo and isolated spirals
in each class and Table~\ref{assign} the class assignment of Virgo galaxies.

\begin{deluxetable}{lccllllccccll}
\tablecaption{Populations of Star Formation Classes for Cluster and
Isolated Environments}
\tablewidth{0pt}
\tablehead{
\colhead{Star Formation Class}&
\colhead{Isolated}&
\colhead{Virgo}}
\startdata
Normal & 83\% (20)  & 37\% (19) \\
Enhanced & 0\% & 6\% (3)\\
Anemic & 4\% (1) & 6\% (3) \\
Truncated/Normal & 8\% (2) & 37\% (19)\\
Truncated/Compact & 4\% (1) & 6\% (3) \\
Truncated/Anemic & 0\% & 8\% (4)\\
Truncated/Enhanced & 0\% & 2\% (1)\\
\hline
Truncated (all) & 12\% (3) & 52\%(27)\\
Anemic (all) & 4\% (1) & 13\% (7) \\
Enhanced (all) & 0\% & \ 8\% (4)
\enddata
\label{tabclass}
\tablecomments{Percentages and numbers 
of isolated and Virgo Cluster galaxies in the
star formation classes described in Section~\ref{sfmorph}. The last
three rows sum the total
numbers of truncated galaxies and the total numbers of galaxies with
anemia or enhancement over at least part of the star-forming disk.}
\end{deluxetable}

\begin{figure}[hbt]
\includegraphics[scale=0.5]{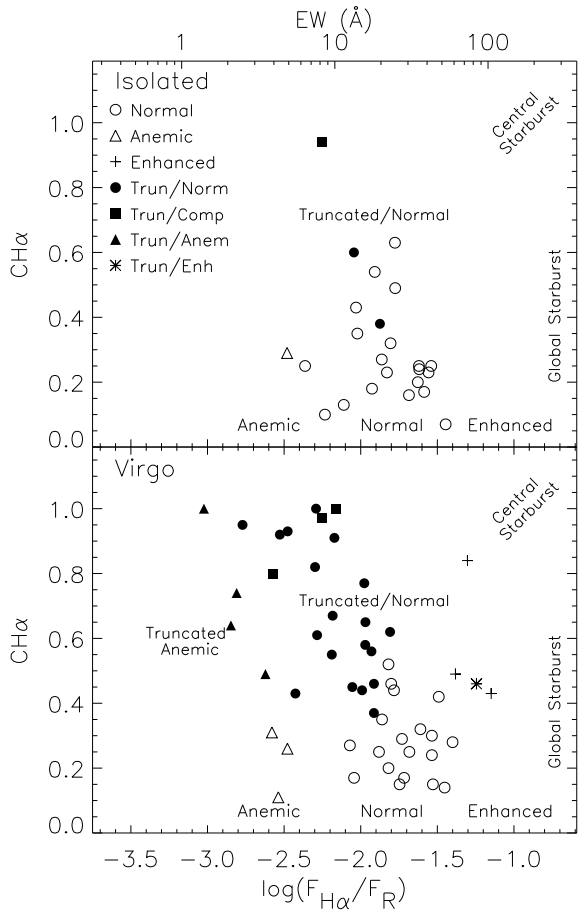}
\caption{H$\alpha$ concentration vs the global NMSFR
for the Virgo sample. The symbols indicate the H$\alpha$
morphology class, which is defined by the radial distribution of star
formation (Section~\ref{sfmorph}).
The upper x-axis provides the equivalent width scale.
The H$\alpha$ morphology classes can be well-discriminated using these
two parameters.
}
\label{hacclass}
\end{figure}

The definition of these classes involves the computation of
NMSFRs over 5 radial bins in a galaxy. 
A second method of discriminating between the different H$\alpha$ morphologies,
which is simpler although less accurate, is based on a plot of the total
NMSFR versus the H$\alpha$ concentration parameter, CH$\alpha$.
Figure~\ref{hacclass} shows that these two parameters 
discriminate fairly well between the classes.
The good separation of classes in the plot
suggests that this type of approach may be useful in identifying
trends in star formation morphologies in large samples of galaxies.
Starburst galaxies would also be well-separated in the plot. As shown in
the figure, central starbursts would have an elevated global NMSFR and a high
H$\alpha$ concentration, since most of the enhanced star formation is occuring
in the center. Global starbursts would have a lower H$\alpha$ concentration, 
because of enhanced star formation throughout the disk.
However, this approach does not work perfectly for all galaxies. 
For example, it does not distinguish between truncated/compact galaxies
and truncated/normal galaxies with severe truncation,
since for these galaxies the H$\alpha$ radial profile differs inside the
radius of 0.3$r_{24}$ which is used in the definition of CH$\alpha$.

\begin{figure*}[hbt]
\includegraphics[scale=0.7,angle=90]{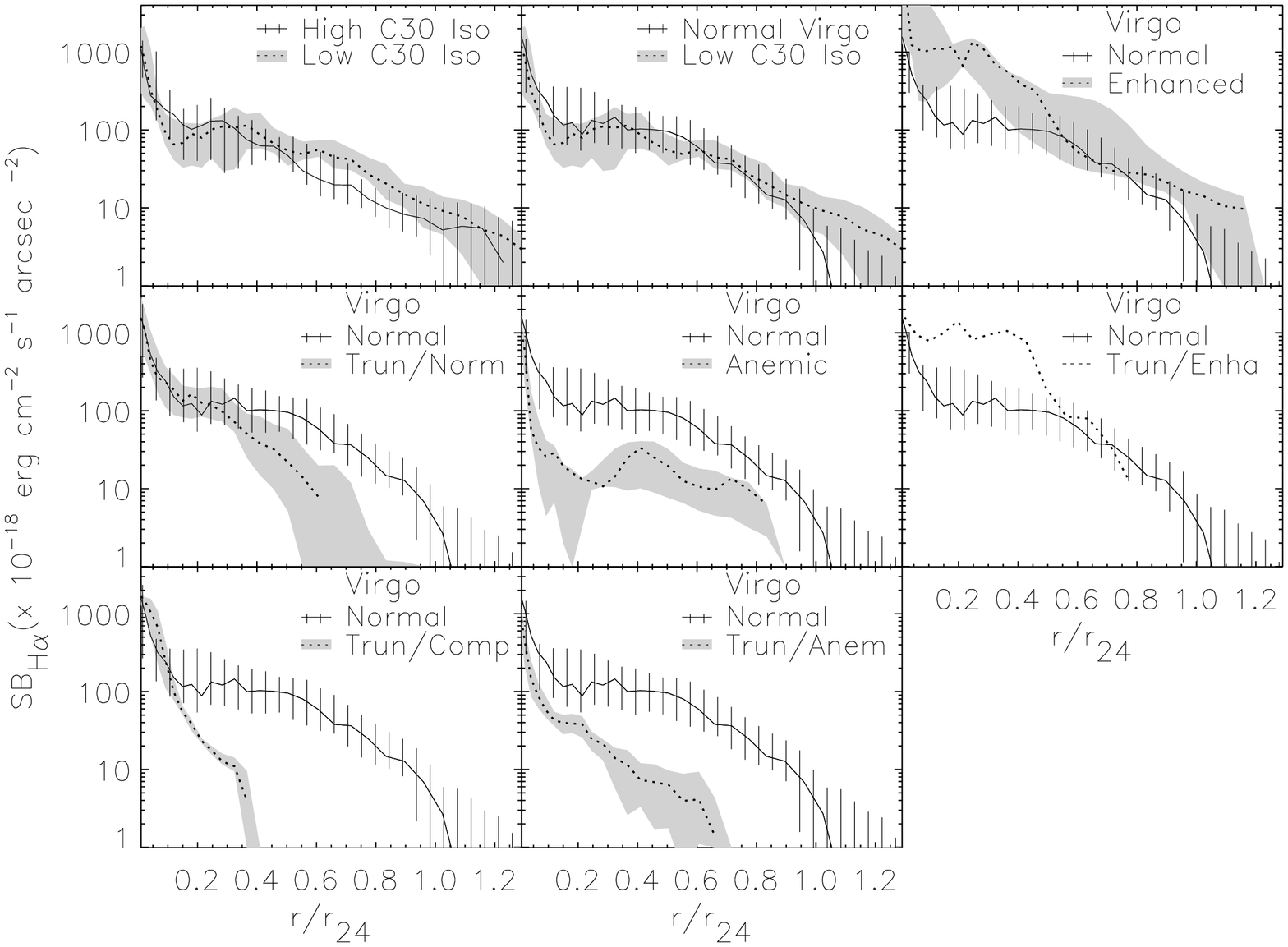}
\caption
{Median H$\alpha$ radial profiles and plotted with interquartile 
range for various classes of galaxies. In the first panel, the 
low ($C30 <$ 0.4) and high concentration ($C30 >$ 0.4) isolated galaxies
are plotted with the indicated shading. The radial distribution of star
formation in isolated galaxies of the two concentration ranges is similar over
much of the star forming disk. In the second panel, the low $C30$ isolated
galaxy median is also similar to the normal Virgo star formation 
morphology class. In the remaining panels, galaxies in 
the normal Virgo spiral class are compared with galaxies in the indicated
Virgo star formation morphology class. 
The normal Virgo spiral median
is significantly different from the median of other star formation
morphologies over much of the disk.}
\label{haclasscomp}
\end{figure*}

The median profiles of the classes are compared in
Figure~\ref{haclasscomp}. (See PIII for an explanation of how
median H$\alpha$ profiles are derived.)
Shading indicates the interquartile range. 
The first panel provides a comparison
between isolated low-$C30$ and high-$C30$ spirals, which are similar
within the interquartile range over most of the disk. In the remaining
panels, the Virgo normal median and range is compared to the  
isolated low-$C30$ and the other classes of Virgo spirals. 
The normal Virgo spiral median
is significantly different from the other classes, except the low-$C30$
isolated spirals (although there is a clear
tendency toward mild truncation at radii beyond $r_{24}$ for the Virgo normal
spirals). 

\begin{figure}[t]
\includegraphics[scale=0.5]{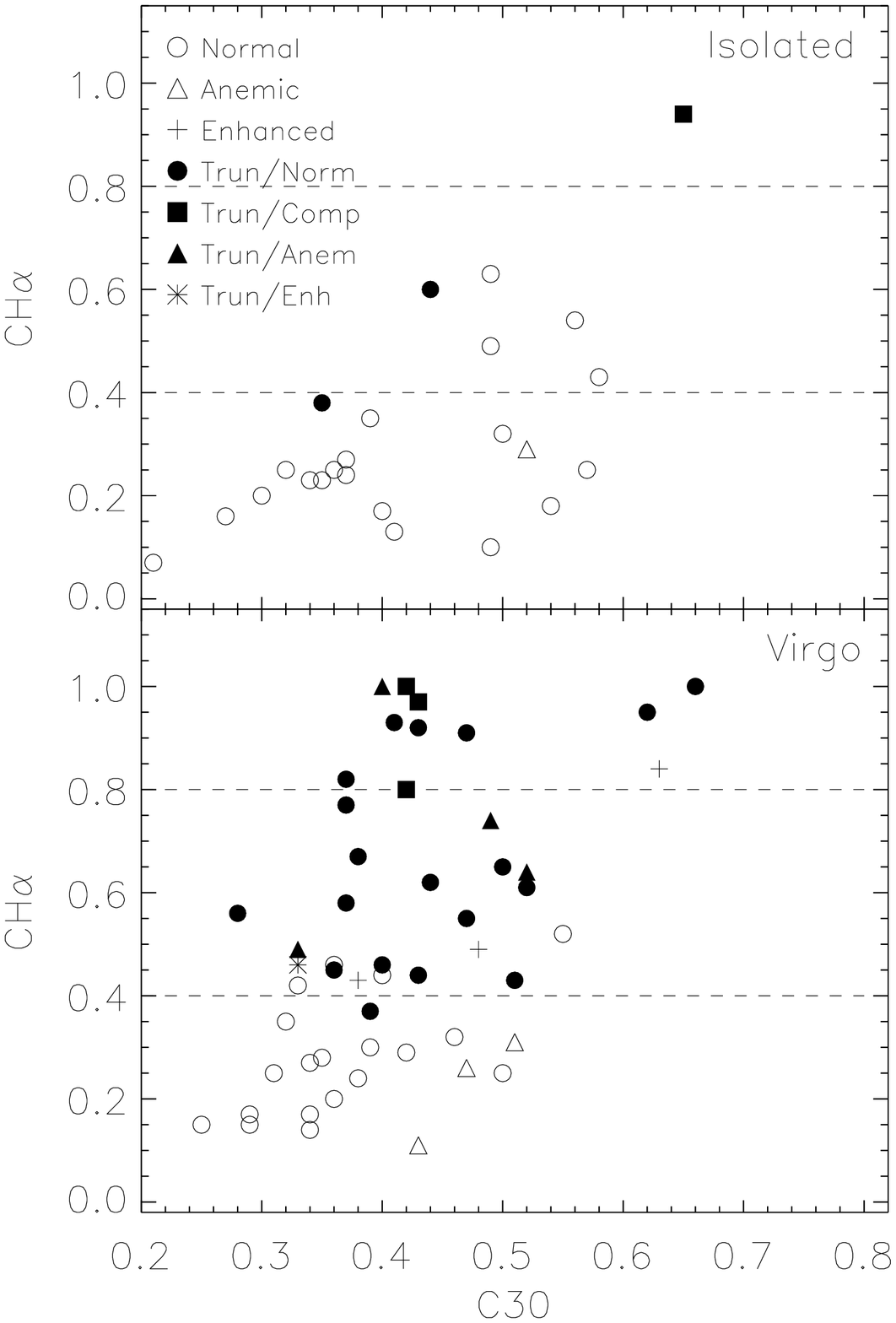}
\caption{Comparison of $C30$ and CH$\alpha$ for the different
star formation classes, showing little correlation between
$C30$ and the star formation morphology.} 
\label{c30haclass}
\end{figure}

There is little correlation with star formation class and $C30$, as shown
in Figure~\ref{c30haclass}. The galaxies in each class show a range
in $C30$. Anemic galaxies show the least range, with all three Virgo anemic
galaxies and the isolated anemic galaxy at higher $C30$ of 0.42 - 0.52 
(see also Bothun \& Sullivan 1980). Two of four
truncated/anemic galaxies also have higher $C30$ values of
about 0.5, but the other two have low $C30$'s of 0.33-0.4. 

The star formation classes are further discussed in 
sections~\ref{normalclass}-~\ref{enhanced}, giving specific galaxy examples.
In Sections~\ref{outarc} and~\ref{pair}, we discuss two other interesting 
subsets of galaxies: galaxies with star formation enhanced locally near
the truncation radius and several apparent pairs in the Virgo Cluster.
In Section~\ref{location}, the locations of galaxies of each class within
the cluster are compared. Section~\ref{himorph} provides a comparison of 
H$\alpha$ and HI morphologies.

\subsection{Virgo Spirals with Normal Star Formation Rates Within $r_{24}$}
\label{normalclass}
About one-third (37\% ; 19 galaxies) of the Virgo Cluster sample spirals have
star formation rates
similar to those of isolated spirals within $r_{24}$.
These galaxies span the full range of $C30$
and are classified later-type spirals (3 Sb and 16 Sc).
Apparently these galaxies have not
experienced a strong and/or recent environmental
interaction.
These galaxies may be recent arrivals in the cluster.
Although these galaxies do not show significant differences in star
formation rates from the isolated spirals, 
some appear to have mildly truncated H$\alpha$ profiles, while others
display other peculiarities. 
For example, NGC 4654 has a peculiar H$\alpha$ distribution
and an HI tail and may be experiencing an ICM-ISM encounter
(Phookun \& Mundy 1995), perhaps in addition to a galaxy-galaxy tidal
encounter (Vollmer 2003).
NGC~4651 appears as a relatively normal Sc within
$r_{24}$, but deep optical images reveal a peculiar linear feature and
several shell-like features  at large radii
(Schneider \& Corbelli 1993; Malin 1994), which may be caused
by a minor merger.
Some of these galaxies also show asymmetric outer arcs of
star formation, possibly indicating ICM pressure (Section~\ref{outarc}).

\subsection{Anemic Virgo Spirals}
\label{anemic}

Anemic galaxies were first defined based on the arm-interarm contrast
in the disk (van den Bergh 1976), which is related 
both to gas content (Elmegreen et al. 2002) and star formation activity.
Historically, the term `anemia' has sometimes been
used to refer to any
galaxy with a reduced global star formation rate.
The spatial resolution
of our study allows us to define anemia more specifically as a
low but measurable H$\alpha$ surface brightness \it across
the disk. \rm
Neither the arm-interarm contrast nor the global star formation rate
distinguish between truncation and anemia.
Three Virgo Cluster galaxies and one isolated galaxy (IC 356)
fall in the anemic class. Images and radial profiles of
the Virgo cluster galaxies are shown in Figure~\ref{anemicfig}.
Four additional Virgo spirals are anemic galaxies which also have
truncated star-forming disks (see Section~\ref{anemtrun}).
The few anemic galaxies tend to fall in the lower left-hand corner of
Figure 1, with low global star formation rates and normal
to low H$\alpha$ concentration.

The dividing line between normal and anemic star formation
is somewhat arbitrary. For
example, if we chose to define anemic galaxies as galaxies which have
NMSFRs reduced by at least three times in all radial bins, the number of
anemics in the Virgo Cluster falls from 3 to 1 and in the isolated sample 
from 1 to 0. The paucity of anemic galaxies shows
that global anemia does not play a dominant role in reduction of total
star formation rates among the Virgo spirals.
While the star formation of some Virgo galaxies may indeed appear
weak across the disk,
isolated galaxies can have equally weak disk star formation (as
also pointed out for a sample of nearby galaxies by Bothun \& Sullivan
1980).

Anemic galaxies should not be confused with low surface brightness
galaxies.
Anemic galaxies have low
H$\alpha$ surface brightness with respect to the continuum, while
low surface brightness galaxies have low continuum and
low-to-intermediate H$\alpha$
surface brightnesses. Because we normalize the H$\alpha$ emission by the R
continuum brightness, the NMSFRs of low surface brightness galaxies
fall in regions of our plots similar those of high surface brightness
galaxies.

\begin{figure*}[t]
\caption{Three Virgo Cluster galaxies with anemic star
formation, i.e., reduced star formation rates across the disk.
In this and the following figures, the R and H$\alpha$ images and
surface photometry profiles are given for each galaxy.
The images are displayed on a log scale, with north up and east to the
left.
The RSA/BST and RC3 morphological types are indicated.
In the surface photometry plots, the R (solid) and
H$\alpha$ (dotted) profiles are plotted as a function of radius in arc
seconds.
The H$\alpha$ profiles were superposed using an arbitrary
zeropoint of 18.945. 
The H$\alpha$ profile is cut at the radius of the outermost HII region.
At upper right the galaxy positions are shown on top of a 
ROSAT x-ray map of the Virgo cluster (B\"{o}hringer et al. 1994).
At lower right, large filled circles identify 
the galaxies in the $C30$-NMSFR plot (see PIII, Figure 8).
In the following figures, additional galaxies within a class are indicated by
smaller fonts in the cluster map and smaller filled circles in the $C30$-NMSFR
plot.
Note how the indicated anemic galaxies fall at low star formation rates
globally, in the inner 30\% of the disk, and in the outer 70\% of
the disk. }
\label{anemicfig}
\end{figure*}

It is illustrative of the past approach of anemic classification to
compare the classifications of van den Bergh (1976) and van den Bergh,
Pierce, \& Tully
(1990). They classify NGC 4548 as anemic, but NGC 4394 as a normal spiral.
They list a number of other galaxies in our sample as anemic, including
(i) NGC 4424, 4457, 4522, 4569, and 4580, which we find to be
severely truncated,
(ii) NGC 4579, and 4689, which are less severely truncated,
(iii) NGC 4293 and NGC 4450 which are truncated/anemic
(iv) NGC 4651, which is normal, and
(v) NGC 4643, 4710, and 4429, which do not have detectable HII regions.
We reemphasize that the term anemic, as used in the past, has been
applied to a variety of star formation distributions.

\subsection{Virgo Spirals with Truncated Star-Forming Disks}
\label{truncated}
Just over half (52\% ; 27 galaxies) of the Virgo sample
show truncation in the star-forming disk within 0.8$r_{24}$.
This class of star-formation morphology is therefore by far
the most common in the Virgo Cluster.
The severity of truncation  varies, 
from moderate truncation in the outer half of
the disk (from 0.4-0.8 $r_{24}$) (16), to severe truncation at radii
less than 0.4 $r_{24}$ (11). 
The star formation morphology within the truncation radius also varies,
from normal (19) to enhanced (1) to anemic (4) to compact (3).

The severely truncated galaxies were referred to as `St' by
Koopmann \& Kenney (1998). In this paper, we give these galaxies a more
detailed designation according to their star formation morphology, which
is either truncated/normal or truncated/compact. 

\subsubsection{Truncated Spirals with Normal Inner Disks}
\label{trunnorm}

The largest subclass of truncated galaxies are those which have normal or
slightly enhanced inner star formation rates compared to isolated spirals.
Eight galaxies are severely truncated, with
normal-to-slightly enhanced (up to a factor of 3.3) star formation
within 0.3-0.4$r_{24}$, but no star formation in the disk
outside this radius.

Three galaxies, NGC 4580, NGC 4405, and IC 3392, 
have severely truncated H$\alpha$ disks with enhanced star formation
at the H$\alpha$ truncation radius, but appear fairly normal inside the
truncation radius. These galaxies were assigned `mixed'
Hubble-type classifications of Sc/Sa and Sc/S0 by Binggeli, Sandage, \& 
Tammann (1985; hereafter BST) because of their peculiar morphology. 
The R images (Figure~\ref{stmix}) show that the stellar disks
beyond the edge of the star-forming disk are mostly regular and
featureless, except for NGC 4580, which has spiral arms. 
NGC 4580 and IC 3392 have symmetric rings
of star formation near the truncation radius, 
containing a large fraction of the H$\alpha$
emission. NGC 4405 has a knottier and asymmetric HII region distribution,
with an arc or partial ring to the southwest. H$\alpha$ major axis
spectra of all three galaxies show regular, rising rotation curves
for the ionized gas (Rubin, Waterman, \& Kenney 1999).
The truncated H$\alpha$ distribution, combined with the regular stellar
isophotes and regular gas kinematics, strongly suggest that these
galaxies are victims of ICM-ISM stripping.

\begin{figure*}[t]
\caption{Three Virgo Cluster galaxies with severely truncated
star-forming disks (within 0.4$r_{24}$). These galaxies
have regular stellar disks and their H$\alpha$ morphologies show
symmetric rings of star formation near the truncation radius. 
These galaxies have likely been stripped by the ICM. 
The galaxies are indicated in the cluster map by large font lettering and
in the $C30$-NMSFR plot by large filled circles. The locations of
other galaxies with severely truncated H$\alpha$ disks are indicated
in the cluster map with smaller font and in the $C30$-NMSFR plot with 
small filled circles. 
See Figure~\ref{anemicfig} for details on the plots.}
\label{stmix}
\end{figure*}

Two severely truncated galaxies
with apparently extraplanar HII regions (NGC~4522 and NGC~4569)
are shown in Figure~\ref{trunnorm1}.
While the stellar disk of the highly inclined NGC 4522 appears
relatively undisturbed, the H$\alpha$ and HI emission 
show a truncated disk
and extraplanar gas reminiscent of a bow shock morphology
(Kenney \& Koopmann 1999; Kenney, van Gorkom, \& Vollmer 2004). 
This strongly suggests that the ISM of NGC~4522 is
being stripped by the gas pressure of the ICM.
NGC 4569 has smooth, relatively featureless, outer spiral arms 
and a ring of star formation at 0.3 $r_{24}$.
A distinct peculiar arm of HII regions, which begins at the
H$\alpha$ truncation radius and has an HI counterpart (Vollmer et al. 2004), 
may be extraplanar (Hensler et al. 2003).
This arm resembles the one-arm gas morphologies commonly seen in the 
ICM-ISM interaction
simulations of Schulz \& Struck (2001) and Vollmer et al. (2001).

NGC 4351, also shown in Figure~\ref{trunnorm1},
has a strongly asymmetric, off-center,
patchy distribution, with a high surface brightness in the circumnuclear 
region. The HI emission is highly asymmetric in the same direction as the 
H$\alpha$ emission (Warmels 1988).
NGC 4351 is one of the galaxies closest to the cluster center,
only 1.7$^{\circ}$ from M87, and it
has a high line-of-sight velocity of 2310 km s$^{-1}$. The location
of the galaxy and the combination of more regular R isophotes and
asymmetric
H$\alpha$ and HI emission make this galaxy a good candidate for an
ongoing ICM-ISM interaction.

The severely truncated galaxy
NGC 4457 (image given in PI) 
has a regular R morphology and a nearly circular outer ring, which
suggests we are viewing the galaxy close to face-on.
The galaxy has a relatively strong but truncated H$\alpha$ disk
with much of the H$\alpha$ emission arising from one peculiar spiral arm.
While this arm may  resemble those seen in ICM-ISM simulations (Schulz
\& Struck 2001),
NGC~4457 is located 9\arcdeg \ from M87 where the ICM is thought to be 
very tenuous.

Of the eight galaxies in the truncated/normal (severe) category,
the amorphous galaxy NGC 4694 seems least consistent with ICM-ISM stripping.
Its H$\alpha$ profile is
borderline truncated/anemic galaxy or truncated/compact.
NGC 4694 features a complex dust morphology (Malkan, Gorjian, \& Tam 1998),
low major axis velocities, (Rubin et al. 1999), and a peculiar
HI distribution, with a small central concentration of HI and a
40 kpc HI tail extending to the neighboring
dwarf irregular galaxy VCC 2062 (van Driel \& van Woerden 1989;
see also Hoffman et al. 1996).
van Driel \& van Woerden (1989) conclude
that an interaction with an intergalactic gas cloud is most likely
the explanation for the structure of NGC 4694, although an interaction
with VCC 2062 or another galaxy is possible.

\begin{figure*}[t]
\caption{Three Virgo Cluster galaxies with star forming disks truncated
within 0.4$r_{24}$, normal-enhanced inner star formation rates, and peculiar
H$\alpha$ distributions, possibly extraplanar emission.
NGC 4569 has a ring of star formation
at about 0.3$r_{24}$ and a detached arm of HII regions, which may be
extraplanar (Hensler et al. 2003).
NGC 4522 appears to interacting with the ICM, based on the
extraplanar HII regions and disturbed HI and radio continuum (Kenney \&
Koopmann 1999; Kenney et al. 2004)
The H$\alpha$ emission in NGC 4351 is offset from the apparent center
in R. 
See Figure~\ref{anemicfig} for details on the plots. }
\label{trunnorm1}
\end{figure*}

The remaining eleven H$\alpha$ truncated galaxies  
have less severely truncated star-forming disks, with truncation
radii of 0.4-0.8$r_{24}$. They also have more HI than
galaxies with severely truncated H$\alpha$ disks, with HI deficiency
parameters
ranging from 0.07-0.81 (HI deficient up to a factor of 7).
Two isolated galaxies, NGC 613 and IC 5273, also have moderately truncated
H$\alpha$ profiles.

\subsubsection{Virgo Spirals with Compact Star-Formation}
\label{truncomp}

Three Virgo Cluster spirals have extremely small H$\alpha$ disks, with
HII complexes distributed asymmetrically and only within the central kpc.
All have $M_B$ between -18 and -19, and all are strongly HI deficient.

These include two 
Virgo galaxies with the most unusual H$\alpha$ morphologies of our sample:
NGC 4424 and NGC 4064, which have compact
H$\alpha$ emission from several circumnuclear HII complexes,
as shown in Figure~\ref{stt}.
The H$\alpha$ surface brightnesses
of these two galaxies within 0.1r$_{24} \sim$ 0.8 kpc are among the
highest in the Virgo and isolated samples (see also Figure~\ref{haclasscomp}),
and most of the H$\alpha$ emission originates in HII complexes with
luminosities of L$_{\rm H\alpha }$=5$\times$10$^{38}$ erg s$^{-1}$.
H$\alpha$ major axis spectra show low line-of-sight velocities
inconsistent with circular motions in the plane suggested by
the outer optical isophotes (Rubin et al. 1999).
NGC 4424 has unusual heart-shaped inner R isophotes and shell-like features
(Kenney et al. 1996) similar to those seen in simulations of mergers 
(e.g., Hernquist
\& Quinn 1988; Hernquist 1992, 1993), suggesting that NGC 4424 is
the product of a recent minor merging event (Kenney et al. 1996).
NGC 4064 has a more regular R morphology
than NGC 4424. It shows very open spiral arms and a bar
in the center. The peculiar dust morphology also suggests a recent minor merger
or tidal interaction (Cortes \& Kenney, in prep.).
Strongly HI-deficient with no gas detected in their outer disks, these
galaxies probably have experienced ICM-ISM stripping in addition to mergers.

The other truncated/compact galaxy NGC~4606 has fainter H$\alpha$ emission,
consisting of $\sim$ 3 faint HII complexes, distributed in a linear feature 
extending from the center toward
the southwest and almost coincident with the major axis. 
A possible companion is NGC~4607 (separation of 4 \arcmin \ =18
kpc and 593 km s$^{-1}$).

The \it isolated \rm severely truncated galaxy, NGC 4984, has a
compact star-forming disk with active star formation distributed
in an incomplete nuclear ring.
The galaxy has a high $C30$ and is
classified as Sa and a starburst (Devereux 1989; Kewley et al. 2001).
Thus severely
truncated H$\alpha$ disks do occur in isolated galaxies, perhaps also
triggered by minor mergers or tidal interactions with low surface
brightness companions.

\begin{figure*}[t]
\caption {Three Truncated/Compact Virgo spirals.
These galaxies have severely truncated star-forming disks. Star formation
is located in several circumnuclear HII complexes, which have a
non-axisymmetric distribution.
The H$\alpha$ surface brightnesses
of  NGC 4424 and NGC 4064 within 0.1r$_{24} \sim$ 0.8 kpc are among the
highest in the Virgo and isolated samples.
The gas in these galaxies has low line-of-sight velocities (Rubin et al.
1999).
NGC 4424 also displays shell-like features and banana-shaped isophotes
(Kenney et al. 1996).
These galaxies are likely to be products of a merger or close tidal
interaction.
See Figure~\ref{anemicfig} for details on the plots. }
\label{stt}
\end{figure*}

\subsubsection{Truncated Spirals with Anemic Inner Disks}
\label{anemtrun}

Four galaxies 
have truncated star-forming disks with anemic star formation within
the truncation radius. All are classified as Sa by BST
and/or by deVaucouleurs (1991; hereafter RC3), although two 
(NGC 4293 and NGC 4380) have $C30$ values similar to isolated Sb or Sc galaxies.
Images of three of them are shown in Figure~\ref{trunanem}.

NGC~4380, NGC 4419, and NGC~4450 appear to have regular stellar morphologies.
NGC 4380 has weak star formation
over much of the disk within the truncation radius of 0.7$r_{24}$,
and a sharp ridge of H$\alpha$ emission on the northwest side
suggestive of ICM pressure.
The highly inclined 
NGC 4419 has a strongly asymmetric CO distribution (Kenney et al. 1990),
suggesting an ongoing ICM-ISM interaction.
In contrast, NGC~4293 has a disturbed stellar morphology suggesting a
tidal disturbance or minor merger.

\begin{figure*}[t]
\caption{Three Truncated/Anemic Virgo Cluster galaxies.
These galaxies have truncated star-forming disks and
anemic inner star formation rates. NGC 4380 and NGC 4293 are classified as
Sa galaxies, but both have a small central R light concentration. The sharp 
ridge of H$\alpha$ emission to the northwest in NGC 4380 is suggestive
of ICM interaction. The stellar morphology of NGC 4293 is disturbed,
possibly due to a tidal interaction.
See Figure~\ref{anemicfig} for details on the plots.}
\label{trunanem}
\end{figure*}

Three additional galaxies, NGC~4579, NGC~4694 and NGC~4772, can be considered
borderline truncated/anemic. All have NMSFRs 
which are low across the disk, but by slightly less than a 
factor of two, and so are classified as truncated/normal. 
The Southern Extension galaxy NGC 4772 
shows kinematic signatures of a past minor merger (Haynes et al. 2000),
and NGC~4694 exhibits peculiarities suggestive of a tidal interaction 
(Section~\ref{trunnorm}).

\subsubsection{Truncated Spirals with Enhanced Total Star Formation Rates}
\label{trunenh}
The only sample galaxy with both enhanced star formation and
a truncated  H$\alpha$ disk is NGC~4299.
Its global H$\alpha$ equivalent width of 84\AA \ is one of the 3 
highest values in the
Virgo sample, and its H$\alpha$ image (Figure~\ref{enhanced3}) 
shows numerous bright HII regions throughout a disk
truncated beyond 0.7$r_{24}$.
The outer extent of the H$\alpha$ distribution is irregular,
except in the southwest where it forms a well-defined ridge
extending over 90\arcdeg, suggesting ongoing ICM pressure.
The most luminous HII complex in the galaxy is located at the
southern end of this ridge.
NGC~4299 is in an apparent pair with NGC~4294,
which is projected 5.8\arcmin \ (27 kpc) away,
and has a similar line-of-sight velocity.
NGC~4294 is similar to NGC~4299 in mass, luminosity, morphology,
HI content, and a relatively high global star formation rate.
Unlike NGC~4299, there is no clear
ridge of HII regions at the outer edge of the H$\alpha$ disk.
The high rates of star formation in these galaxies might be
due a tidal interaction, perhaps enhanced in NGC~4299 due to
an ongoing ICM-ISM interaction.
The location of this pair, only 2.4$^{\circ}$ from M87,
makes an ICM-ISM interaction plausible.

\subsection{Virgo galaxies with enhanced inner and total star formation
rates}
\label{enhanced}

Six percent (3 galaxies) of the Virgo sample has total and inner
star formation rates much higher than those in the isolated sample,
and one additional galaxy (discussed separately in Section ~\ref{trunenh})
has an enhanced inner disk combined with a truncated 
outer disk. 
Three of these enhanced 
galaxies are intermediate mass galaxies with $M_B$ between
-19 and -18, and one is massive, with $M_B$=-21.

\begin{figure*}[t]
\caption{Three Virgo Cluster galaxies 
with enhanced or truncated/enhanced star formation. There
is evidence for tidal forces and/or gas accretion in all three cases.
The truncated/enhanced galaxy
NGC 4299 may plausibly be experiencing an ICM-ISM interaction. 
In the $C30$-NMSFR plot, NGC 4299 is indicated by a filled square.
Note the relatively high position of all of the enhanced galaxies in
the $C30$-NMSFR diagram and that these galaxies tend to be located
in the cluster outskirts. All of these galaxies except NGC~4303 
have $M_B$ between -18 and -19.
See Figure~\ref{anemicfig} for details on the plots. } 
\label{enhanced3}
\end{figure*}

The 3 Virgo sample galaxies with the highest normalized
massive star formation rates are NGC~4383, NGC~4532, and the Truncated/Enhanced
galaxy NGC~4299. (Figure~\ref{enhanced3})
The normalized star formation rates of spiral galaxies are a function of
both luminosity and morphological type (Boselli et al. 2001),
and these 3 galaxies
have total NMSFRs which are  2.5-5 times higher than the isolated median, and
2-3 times larger than any isolated sample
galaxy of similar luminosity or central concentration (PIII).
Compared to the isolated galaxies,
the inner NMSFRs of these galaxies are even more extreme than the total NMSFRs,
with values 3-4 times larger than any isolated sample galaxy
(see Figures 4 and 8 of Paper III).
The total H$\alpha$ equivalent widths of these 3 galaxies are 70-100\AA,
which is somewhat higher than those in the nearby, well-known starburst
galaxies M82 and NGC~1569 (Kennicutt \& Kent 1983).
All have equivalent widths which are on the high end of the
distribution shown by Boselli et al. (2001) for galaxies in many
environments,
for the appropriate galaxy luminosity.
Nearly all non-cluster spiral galaxies with H$\alpha$ equivalent widths
above 60\AA \ are strongly disturbed
or in galaxy pairs (Kennicutt et al. 1987).

The Sm/Im galaxy NGC~4532 has the highest NMSFR in the Virgo sample, 
with an  H$\alpha$ equivalent width of 88\AA.
Its H$\alpha$ image (Figure~\ref{enhanced3}) shows vigorous star formation 
throughout the galaxy, with no sharp outer boundary in H$\alpha$ 
that would indicate ICM pressure.
The galaxy is moderately HI-rich (HI def=-0.35),
and both NGC~4532 and the nearby DDO137 are associated with a large
extended, bi-lobal, HI cloud (Hoffman et al. 1993).
Much of the HI is concentrated around the 2 optical galaxies, but
about one-third seems kinematically
and spatially distinct from the individual galaxies
(Hoffman et al. 1999).
The enhanced star formation in NGC~4532 might therefore be caused by 
tidal forces and/or gas accretion.

NGC~4383 is an amorphous (BST) galaxy with a 
large bulge-to-disk ratio ($C30$=0.63).
Most of the isolated galaxies of similar concentration are
S0 galaxies without significant HII regions.
The H$\alpha$ image shows a biconical filamentary structure strongly
suggestive of a starburst outflow, numerous HII regions in the
central 2 kpc, and an irregular string of bright  HII regions
extending into the outer galaxy.
The  dS0 galaxy UGC~7504 (VCC 794),
3.3 magnitudes fainter than NGC~4383,
is very nearby in projection (2.5\arcmin \ = 12 kpc), with
a line-of-sight velocity difference of 800 km s$^{-1}$, and
no H$\alpha$ emission.
NGC~4383 is HI rich (HI def = -0.53) and has no sharp outer boundary in 
H$\alpha$, suggesting that a process other than ICM pressure,
probably a tidal interaction and/or gas accretion,
is responsible for its enhanced star formation.

The third enhanced Virgo galaxy, 
NGC~4303, is one of the largest spirals in Virgo ($M_B$=-21). It
has an H$\alpha$ equivalent width of 61\AA, \ which
is unusually high for a large mass spiral (Boselli et al. 2001). 

The Virgo galaxies with the highest NMSFRs are all HI normal to HI rich,
and preferentially H$\alpha$-enhanced in the inner galaxy.
All have apparent nearby companions, which suggests that
low-velocity tidal interactions may play a role in the
enhancement in star formation rates, 
as in observed in galaxy pairs outside of clusters (Kennicutt et al. 1987).
At least one of them is associated with a
large, extended cloud of HI, suggesting that HI accretion,
perhaps combined with tidal forces,
plays a role in the enhanced star formation rates.

There is evidence that ICM-ISM interactions enhance the NMSFR
of some Virgo galaxies, but with the possible exception of 
NGC~4299 (Section ~\ref{trunenh}),
the enhancements appear to be local and globally modest.
As discussed in Section~\ref{outarc},
there are a number of other Virgo galaxies whose
H$\alpha$ morphologies are strongly suggestive of locally enhanced NMSFR due to
ICM pressure.

\subsection{Galaxies with Enhanced Outer Disks?}
\label{outarc}
One might expect
an additional class of galaxies with significantly
enhanced outer disks. Such a morphology might arise
from ICM induced star formation in a galaxy.
However, only 3 Virgo Cluster galaxies  (NGC 4519, NGC 4713, NGC 4532),
show evidence for outer disk star formation rates
which are enhanced above levels seen in isolated galaxies,
and these galaxies also have high NMSFRs in their inner disks.
As discussed in Section~\ref{enhanced}, the star formation
properties of these galaxies are most likely influenced by tidal
interactions or gas accretion.

Seven spirals show asymmetric enhancements
in star formation at the outer edge of the H$\alpha$ disk.
These enhancements are modest, since in none of the galaxies do the
NMSFRs measured in
any radial bin exceed the values found for isolated spirals (see Figure
8 in Paper III).
The H$\alpha$ distributions in these galaxies range from severely
truncated (NGC~4405, Figure~\ref{stmix}),
to normal (NGC 4178, NGC 4189, and NGC 4654, Figure~\ref{edge}),
and 1 is truncated/anemic (NGC~4380, Figure~\ref{trunanem}).
Two are in apparent close pairs (NGC 4298 and NGC 4647, Figure~\ref{pairfig}),
and a third may also be in a binary pair (NGC~4654, Vollmer 2003),
and in these cases the roles of tidal and ICM effects are unclear.
For the 4 systems not in pairs, the H$\alpha$ morphologies are
just what is expected from ongoing ICM pressure.
These asymmetric enhancements are different from the
axisymmetrically enhanced outer  H$\alpha$ rings in the
H$\alpha$-truncated  galaxies  NGC 4580 and IC 3392 (Section~\ref{trunnorm}).
Guided by the simulations of Schulz \& Struck (2001) and Vollmer et al.
(2001),
we suggest that those galaxies with asymmetrically enhanced outer
H$\alpha$ edges are
presently experiencing ICM pressure, whereas those with
symmetric outer rings may be observed after peak pressure,
when the truncated gas disks `anneals' (Schulz \& Struck 2001).

\begin{figure*}[t]
\caption{Three Virgo Cluster galaxies which show locally enhanced
H$\alpha$ emission in asymmetric arcs near the edge of the star-forming
disk.
NGC 4380 (Figure~\ref{trunanem}) and
pair galaxies NGC 4298 and NGC 4647 (Figure~\ref{pairfig})
also show similar outer arcs of star formation. However this local
enhancement is not significant enough to increase the NMSFR over the
whole radial bin, as shown at lower left. (See Figure~\ref{anemicfig} for 
details on the plot, but note that, unlike preceding
figures, the plot at lower right depicts the NMSFRs within three
smaller radial bins: 0.3$r_{24} < r  < $0.5$r_{24}$, 0.5$r_{24} < r  < 
$0.7$r_{24}$, and 0.7$r_{24} < r < $1.0$r_{24}$.) }
\label{edge}
\end{figure*}

\subsection{Pairs in the Virgo Cluster}
\label{pair}

At the high relative velocities characteristic 
of many tidal encounters in clusters,
the interacting partners move away from each other so quickly
that the occurrence of a tidal interaction, the participants, and
the interaction parameters are hard to determine.
One category of tidal interaction where the interacting galaxies
can be identified is bound pairs.

We have analyzed the magnitudes, velocities, and positions of galaxies  in
the Virgo Cluster Catalog (BST)
to assess the significance of apparent pairs in the cluster.
We have identified apparent pairs which
include at least one galaxy brighter than B=14 ($M_B$ = -17),
are similar in mass with a blue magnitude difference of less than 3,
and are close together on the sky, with a position difference of less than
10\arcmin (= 45 kpc projected).
There are 61 such pairs in Virgo, and 49 of these have
measured velocities for both galaxies, according to BST.
The probability that
the apparent pairs are due to chance superpositions along the line-of-sight
can be estimated by assuming a Maxwellian distribution of
velocity differences (e.g., Binney \& Tremaine 1987).
For a cluster velocity dispersion of $\sigma$=700 km s$^{-1}$,
one would expect only 2 pairs
with velocity differences less than 200 km s$^{-1}$, whereas 16 such
low-velocity pairs exist. This clear excess
implies that most of these low-velocity apparent pairs are true pairs.

Four of these low velocity pairs are in our Virgo sample, and
three are shown in Figure~\ref{pairfig}.
All of these galaxies have at least mildly truncated H$\alpha$ disks.
Two have asymmetric H$\alpha$ enhancements at the truncation radius,
on the side closest to the companion, as well as asymmetric R morphologies
(NGC 4298 and NGC 4647).
Two others, NGC 4294/4299 (Figure~\ref{enhanced3})
have mildly to strongly enhanced H$\alpha$ equivalent
widths, similar to the enhancements observed in some non-cluster pairs  
(Kennicutt et al. 1987).

Since tidal interactions affect both stars and gas, it is likely that
ICM-ISM interactions and not tidal interactions are responsible for
the truncated gas disks + asymmetric stellar morphologies 
in these Virgo pair galaxies. This is consistent
with the location of these pairs, all of which are projected within  
3.2\arcdeg \ from M87.
Thus, these galaxies have likely experienced both ICM-ISM and tidal
interactions.
The presence of these pairs, but the absence of recent major mergers
in the Virgo Cluster suggests
that the most of the pairs may be disrupted by tidal
interactions with other galaxies or the cluster before they can merge.

\begin{figure*}[t]
\caption{Three apparent pairs in the Virgo Cluster.
These galaxies appear close in projection
and have similar line-of-sight velocities. Another apparent pair is
NGC 4294/4299 (see Figure~\ref{enhanced3}). Note that a few galaxies
are enhanced globally (NGC 4294 and NGC 4299) and all have substantial
inner star formation. 
See Figure~\ref{anemicfig} for details on the plots. }
\label{pairfig}
\end{figure*}

\subsection{Location of Galaxies by Class in the Cluster}
\label{location}

The locations of the different H$\alpha$ classes 
on a ROSAT x-ray map of the Virgo cluster (B\"{o}hringer et al. 1994)
are shown in the upper right
panels of Figures~\ref{anemicfig}-~\ref{pairfig} and Figure~\ref{rosatmaps}. 
Figure~\ref{radialtype} shows the radial distribution in the cluster
of each H$\alpha$ type. For this plot, we have divided the Virgo sample 
into 3 radial bins:
$\leq$3\arcdeg, 3\arcdeg-4.5\arcdeg, and $\geq$4.5\arcdeg,
with about one-third the total sample in each.

Inside 3\arcdeg, $\sim$80\% of the galaxies are truncated in some way,
and only $\sim$10\% of the galaxies are normal.
In the inner 2\arcdeg, all 7 of the sample galaxies are truncated.
The fraction of galaxies which are truncated in some way
decreases strongly with radius, to 50\% in the mid-cluster, 
and 30\% in the outer cluster (see also Dale et al. 2001). 

Normal H$\alpha$ types also vary strongly with radius.
There are no normal galaxies within 2.4\arcdeg of M87,
but beyond 3\arcdeg about half of the galaxies are H$\alpha$-normal.
Several of the `H$\alpha$-normal' galaxies  
in the intermediate radial bin actually
seem to be H$\alpha$-truncated very far out in their disks.

The anemic, enhanced, truncated/compact, and truncated/anemic 
radial distributions are all roughly flat, within the errors,
although there are no more than 2 galaxies in any radial bin.
The outer cluster ($\geq$4.5\arcdeg) 
contains half the anemic and enhanced galaxies,
but only 15\% of the truncated/normal galaxies.

While truncated galaxies are clearly concentrated toward the center,
some are found far out.
Some could have been stripped in the past
and carried on highly radial orbits into the outer cluster.
Mamon et al (2004) find that the maximum radius reached by a galaxy which
has passed through the inner cluster to be 1-2.5 virial radii, which in
Virgo is about 1.7-4.1 Mpc. The outer cluster truncated/normal galaxies 
are well within this limit.
Some outer cluster truncated galaxies
appear more peculiar than simply ISM-stripped spirals 
(e.g., NGC~4694, NGC~4772; see Section~\ref{trunnorm}),
and their ISM properties may not be due to stripping in a core passage
(see also Sanchis, Lokas, \& Mamon 2004).

The lower right panel of Figure~\ref{rosatmaps} shows the positions of the 
nine candidates for systems presently experiencing ICM pressure, including
extraplanar HII regions (Section~\ref{trunnorm}) or asymmetrically enhanced 
outer H$\alpha$ arcs (Section~\ref{edge}). While, this 
sample of galaxies is not chosen as objectively as our H$\alpha$ classes,
it does give some 
indication of where ICM pressure may be affecting the galaxies.
The figure shows evidence for active ICM pressure out as far as
about 4.5\arcdeg \ from M87, or 1 Abell radius, which 
is beyond the x-ray contours shown in Figure~\ref{rosatmaps}.
ROSAT has detected x-ray emission further out than is shown in 
Figure~\ref{rosatmaps},
although it is hard to know its precise extent 
due to confusion with galactic emission (B\"{o}hringer et al. 1994).

These results suggest that 
anemia and starbursts are caused by 
mechanisms which operate throughout the cluster, even in the outskirts,
such as galaxy-galaxy tidal interactions.
Truncation is caused by a
mechanism which operates most effectively in the cluster center, 
such as ICM-ISM stripping.

\begin{figure}
\caption{Cluster locations of galaxies of the indicated H$\alpha$ classes
on a ROSAT x-ray map of the Virgo cluster (B\"{o}hringer et al. 1994).
The x-ray peak coincides with the giant elliptical M87,
and secondary peaks correspond to
M86 (1.5\arcdeg W of M87), and M49 (4.5\arcdeg W of M87),
each of which are associated with sub-clusters.
Similar maps for other classes are found in the upper right panel of
Figures~\ref{anemicfig}-~\ref{pairfig}. No normal galaxies are found within 
2.4\arcdeg of M87, while truncated/normal galaxies are typically found closer
to the cluster center. Anemic galaxies are found throughout the cluster.
Galaxies which have extraplanar HII regions or asymmetric 
H$\alpha$  enhancements at the outer edge of a truncated H$\alpha$ disk (shown 
in plot at lower right) are located up to 4.5\arcdeg from M87.
Such galaxies may be presently experiencing ICM pressure.}
 \label{rosatmaps}
\end{figure}

\begin{figure}[t]
\includegraphics[scale=0.6,angle=-90]{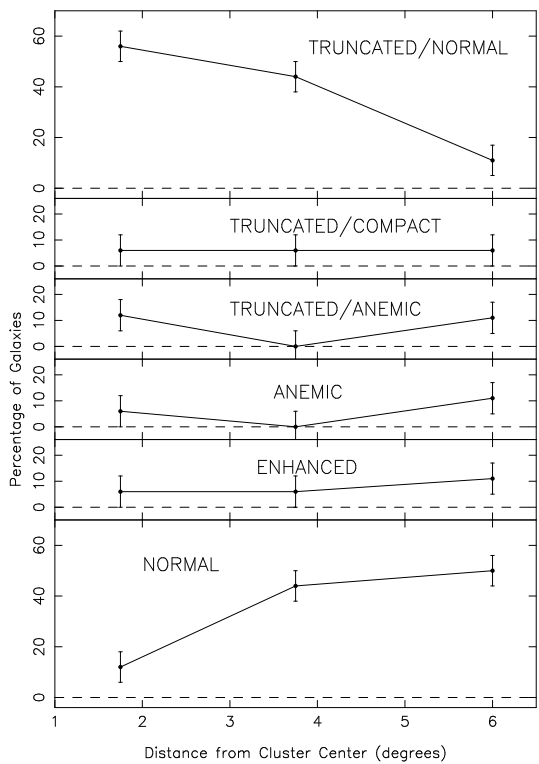}
\caption{
Cluster radial distributions of each H$\alpha$ class, plotted as percentages of
all the galaxies in that radial bin. 
Bins are $\leq$3\arcdeg, 3\arcdeg-4.5\arcdeg, and $\geq$4.5\arcdeg.
Error bars correspond to 1 galaxy, comparable to the uncertainty in
our H$\alpha$ classification. 
There is a clear radial dependence for the normal and truncated/normal
classes, with fewer normal and more truncated/normal galaxies closer to
the cluster center. Other H$\alpha$ classes have a flatter distribution (but
contain fewer sample galaxies). }
\label{radialtype}
\end{figure}

\subsection{H$\alpha$ and HI Morphologies of Sample Galaxies}
\label{himorph}

\begin{figure*}[t]
\includegraphics[scale=0.9]{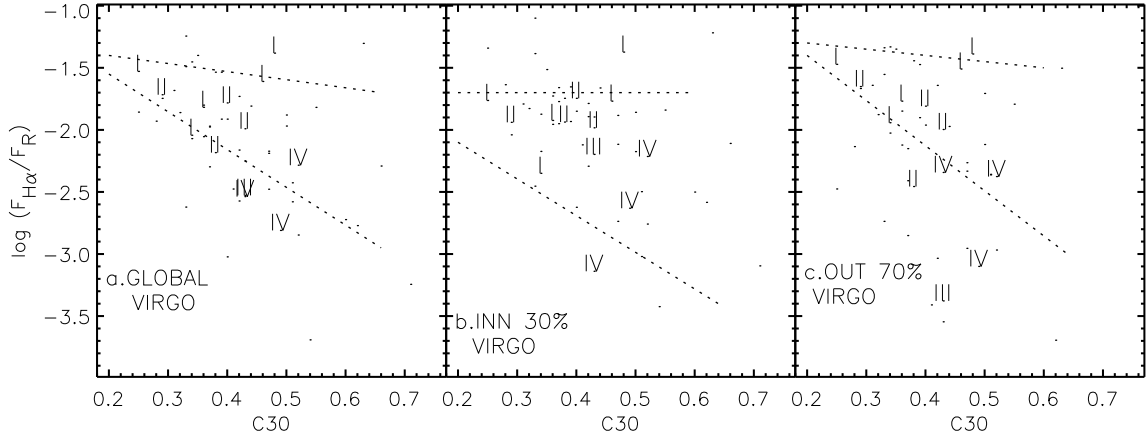}
\caption{NMSFRs versus central concentration parameter $C30$
for (a) the whole galaxy, (b) the inner 30\% of the optical disk, and
(c) the outer 70\% of the optical disk, for the Virgo sample.
The Roman numerals indicate the
HI morphologies described by Cayatte et al. (1994).
There is a clear trend between HI
radial morphology and star formation rates.}
\label{fccay}
\end{figure*}

In Paper III we find 
a strong correlation between the H$\alpha$-based NMSFR and the HI deficiency
parameter. In this section, we compare the H$\alpha$ morphologies 
with HI radial morphologies. Cayatte et al. (1994) describe 4 types 
of HI profiles for 17 Virgo Cluster
spirals: Group I galaxies have normal HI distribution over most of
disk compared to field spirals, Group II galaxies
have lower inner HI densities and somewhat truncated outer distributions,
Group III galaxies have severely truncated HI distributions, but similar
central surface density to field spirals, and Group IV galaxies have a
low surface density of HI as well as a central HI hole.
We have eleven galaxies in common with the sample of Cayatte et al.
Figure~\ref{fccay} shows the 
HI group designation of these galaxies in the $C30-F(H\alpha)/F(R)$ plot.
There is a clear correlation, although not a perfect one-to-one correspondence,
between the HI and H$\alpha$ radial distributions.
The galaxies with severely truncated HI disks (III) also have 
the most truncated H$\alpha$ disks.
The most anemic galaxies in H$\alpha$ are generally those with low HI surface
densities across the disk (IV).
Galaxies with more peculiar HI morphologies also tend to have peculiar
H$\alpha$ morphologies.

\begin{figure}[t]
\includegraphics[scale=0.5]{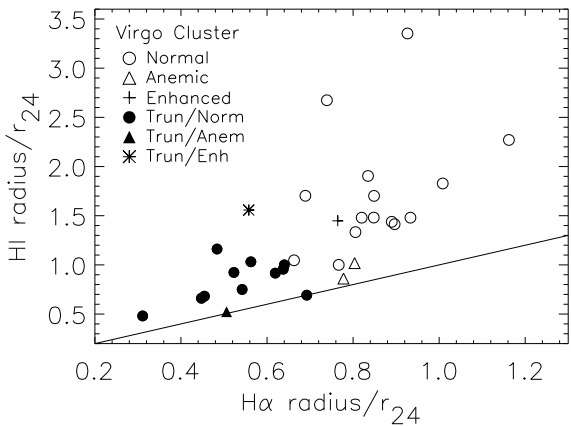}
\caption{Comparison of H$\alpha$ radii to HI radii for galaxies
with HI maps (Warmels 1988; Cayatte et al. 1994; Kenney et al. 2004; Crowl
2004, in prep.). The symbols indicate the H$\alpha$ morphology classes.
There is a good correlation between H$\alpha$ and HI radii. In particular,
galaxies with truncated H$\alpha$ disks also have truncated HI disks.}
\label{hahirad}
\end{figure}

Figure~\ref{hahirad} compares HI and H$\alpha$ radii, normalized by
optical radii, for the Virgo sample. The HI radii are derived from
HI maps by Warmels (1988), Cayatte et al. (1994), Kenney et al. (2004)
and Crowl (2004, in prep.).
The H$\alpha$ radius is defined as the radius containing 95\% of the 
H$\alpha$ emission ($r_{H\alpha95}$ in PI). 
There is a good correlation between the HI and H$\alpha$ radii,
with HI radii typically 0-100\% larger than H$\alpha$ radii. 
Galaxies with H$\alpha$-truncated disks also have HI-truncated disks,
although the HI radii are somewhat larger on average.
This may be partly because star formation is inhibited or very inefficient
in the outermost gas of truncated spirals, just as it is in regular spirals,
due, for example, to an insufficient threshold density for star formation
(Kennicutt 1989).
But it is also partly due to low resolution (typically 30-45\arcsec) for most 
of the HI
measurements, which can result in an overestimate of the true HI diameter.
For those few truncated galaxies with higher ($\sim$15\arcsec) resolution HI
observations, the radii ratios are closer to 1.

The good overall agreement between HI and H$\alpha$ distributions indicates
that H$\alpha$ is a reasonable tracer of the environmental effects on the ISM.
In particular, the H$\alpha$-truncated galaxies all have truncated HI 
distributions, with HI diameters similar to or slightly larger than their
H$\alpha$ diameters. This can be understood as a natural outcome of the
requirement of a threshold density for star formation (Kennicutt 1989).
It also indicates that H$\alpha$ truncation is due to a
truncated gas disk, rather than a gas disk which is simply not forming stars,
due to an increased threshold density for star formation.

\section{Environmental Effects in the Virgo Cluster}

In the preceding sections, we began to link 
the H$\alpha$ morphologies of cluster galaxies
with the environmental effects experienced by them.
In this section we further establish the links, 
and discuss the implications of our results for
cluster galaxy evolution.

\subsection{ICM-ISM Interactions}

The truncation of the star-forming disks in the majority of Virgo
Cluster spirals,
particularly those with regular stellar isophotes, is strong
evidence that ICM-ISM interactions are significant in the evolution
of Virgo Cluster galaxies.
The strong central concentration in the cluster of truncated galaxies
supports the association of truncation with ISM stripping by the 
ICM.
The normal to slightly-enhanced star formation rates in
the inner galaxy disks imply that ICM-ISM interactions have
not had a large effect on the inner regions of most Virgo spiral galaxies.
Modest local H$\alpha$ enhancements in some of the Virgo sample galaxies
are strongly suggestive of ICM pressure (Section~\ref{outarc}),
but those Virgo galaxies with the largest global and inner
star formation rate enhancements appear
to be more strongly affected by processes other than ICM pressure.
While large enhancements in star formation may be
caused by ICM-ISM interactions in some clusters 
(e.g., Dressler \& Gunn 1983; Gavazzi et al. 2001),
in Virgo the enhancements due to ICM pressure seem to be modest and local.

Truncated gas disks in galaxies with relatively undisturbed stellar disks
are predicted from simple physical considerations of ICM-ISM interactions 
(Gunn \& Gott 1972),
and are seen in 3D simulations
(Abadi, Moore, \& Bower 1999; 
Quilis, Moore, \& Bower 2000; Schulz \& Struck 2001;
Vollmer et al 2001).
The key parameters of an ICM-ISM interaction are the ICM pressure, which
in turn depends on the ICM density and
the velocity of the galaxy with respect to the ICM,
and the disk angle with respect to the ICM wind direction.
All of these vary as a galaxy follows its orbital path within the
cluster, and the present state of a galaxy depends on this
orbital history. Of particular importance is the time since (until)
maximum ram pressure.
Since many HI-deficient galaxies in clusters have highly radial
orbits (Dressler 1986; Solanes et al. 2001),
the ram pressure can vary by factors of 20-100
during the orbit, and gas may even fall back into
the galaxy after being pushed outward (Vollmer et al. 2001).
The interaction also depends on the properties of the galaxy
and its ISM, particularly the total mass distribution and
the ISM mass distribution.
In addition, what we observe depends on the viewing angle
between the vector of the galaxy's motion through the ICM
and the line-of-sight.
The oldest galaxies in the cluster have made 5-10 orbits,
although each orbit will be different in the changing potential
of the accreting, growing, lumpy, unrelaxed cluster.
A challenge is to recognize the effects of these various
parameters in particular galaxies, and to identify the evolutionary stages
of the interactions.

Some progress has been made distinguishing active from past stripping.
Simulations indicate that truncated gas disks can survive for much more than 
10$^8$ yr (Schulz \& Struck 2001), 
so some of the Virgo spirals with truncated gas disks were likely stripped
a long time ago.
However, galaxies with extraplanar HII regions or asymmetric H$\alpha$ 
enhancements at the outer edges of truncated gas disks
are likely experiencing active ICM pressure (Section~\ref{outarc}).
Among the best examples of active ICM-ISM stripping in the Virgo Cluster
is NGC 4522, a highly inclined spiral with a normal stellar disk,
truncated HI, H$\alpha$ and radio continuum disks,
and extraplanar emission from all 3 ISM tracers displaced to one side of the 
disk (Kenney \& Koopmann 1999; Kenney et al. 2004).
The opposite, leading edge of the galaxy has enhanced polarized radio emission,
and the flattest spectral index, suggesting ongoing ICM pressure
(Vollmer et al. 2004).
Apart from the extraplanar HII regions, the truncated 
H$\alpha$ morphology of NGC~4522 resembles the majority of Virgo spirals.

The combination of a galaxy's rotation with its forward motion
relative to the ICM create an asymmetric ram pressure (Kritsuk 1983).
Simulations including the effects of rotation on ICM-ISM interactions
show in some phases one dominant extraplanar gas arm emerging from
the outer edges of truncated gas disk 
(Vollmer et al. 2001; Schulz \& Struck 2001).
Such arms may be observed in NGC~4548 (Vollmer et al. 2000),
NGC~4654  (Phookun \& Mundy 1995; Vollmer 2003),
and NGC~4569 (Vollmer et al. 2004; Hensler et al. 2003; Kenney et al. in prep).
Simulations show that
the arms can transfer angular momentum from the inner gas disk,
which can make a contracted, denser, `annealed' inner gas disk, 
which has some resistance to further stripping (Schulz \& Struck 2001).
A couple of the Virgo galaxies (NGC~4580, IC~3392; see Figure~\ref{stmix})
show rings of enhanced star formation at the truncation radius,
which resemble the annealed gas rings of Schulz \& Struck (2001).

\subsection{Galaxy-Galaxy Interactions}

Gravitational interactions between galaxies
vary from mergers and accretions to penetrating collisions
to non-penetrating encounters. The variety in collision parameters,
particularly mass ratio, impact parameter, and relative velocity,
results in a range of effects on the colliding galaxies.
Non-merging tidal encounters are common in clusters, and multiple,
fast tidal encounters may strongly influence a galaxy's morphology over
a Hubble time (Miller 1988; Moore et al. 1998).
Simulations of merging galaxies of similar mass
show that extensive tidal tails containing stars and
gas are produced (e.g., Toomre \& Toomre 1972; see also Hibbard 1995).
Deep imaging of Virgo Cluster galaxies shows some
galaxies with low surface brightness tails (Malin 1993),
which have likely been caused by gravitational interactions.
Since tails in clusters are quickly destroyed by cluster tidal fields 
(Mihos 2004),
the Virgo galaxies with observed tails are a small fraction of the 
cluster merger and collision populations.
A likely example of a recent, high-velocity tidal encounter is the
disturbed Virgo Cluster galaxy NGC 4438, which may have experienced a
direct collision (Kenney et al. 1995; Kenney \& Yale 2002).
The H$\alpha$ emission is dominated by a peculiar
one-sided filamentary H$\alpha$ nebula associated with the interaction,
although the global star formation rate is modest.

There is evidence for major and intermediate-mass ratio mergers 
in many Virgo galaxies, including the ellipticals M87, M49, and NGC 4365 
(Gavazzi et al. 2000; Lee, Kim, \& Geisler 1997; Davies et al. 2001),
the S0's NGC~4550, NGC 4382, NGC~4262, NGC~4643 
(Rubin, Graham, \& Kenney 1992; Rix et al 1992; Schweizer \& Seitzer 1988;
van Driel and van Woerden 1991; Richter, Sackett, \& Sparke 1994),
and the Sa's NGC~4698 and NGC~4772 (Bertola et al. 1999; Haynes et al. 2000).
Most of these mergers probably occurred $>$1 Gyr ago, since the galaxies
do not appear very disturbed.
There are no obvious candidates for an ongoing major merger in
the Virgo Cluster; however, there is at least one clear case in the Coma
Cluster (Bravo-Alfaro et al. 2001) and
merger galaxies are frequently seen in higher-z clusters
(van Dokkum et al. 1999).
There are clear examples of recent minor mergers, even among the Sc galaxies.
The Virgo spiral NGC~4651, which appears as a relatively normal Sc within
$r_{24}$, has a peculiar 
linear feature and several  low surface brightness shell-like features
at large radii (Schneider \& Corbelli 1993; Malin 1994),
suggesting a recent minor merger.
Its star formation rate and distribution are not that peculiar, 
and its H$\alpha$ class is normal.
Thus H$\alpha$ is not an especially good tracer of old mergers 
or recent minor mergers
although we note that NGC~4698 is anemic and NGC~4772 is borderline
truncated/anemic,
so anemia may result from some types of mergers.

In our Virgo survey,
H$\alpha$ distributions classified as  enhanced or truncated/compact
are likely caused by tidal interactions, including minor mergers.
The H$\alpha$ morphologies of the 3 Virgo galaxies
with the largest star formation rates, NGC~4299, NGC~4383, and NGC~4532
seem consistent with those seen in non-cluster tidally interacting systems.
In these systems, enhancements are observed in both the global and nuclear 
NMSFRs, 
but the enhancements are generally greater near the center (Kennicutt 1998).
The inward gas flow and strongly enhanced circumnuclear star formation
in the truncated/compact galaxies NGC~4424
(Kenney et al. 1996) and NGC~4064 (Cortes \& Kenney, in prep.)
are likely due to recent minor mergers.
Tidal interactions are expected to occur throughout the cluster,
including the outskirts, where lumpy infall is occurring.
This is consistent with the locations
of galaxies with enhanced and truncated/compact H$\alpha$ profiles,
which are found throughout the cluster.

These H$\alpha$-selected tidally interacting galaxies are only a 
fraction of all tidally interacting galaxies.
Rubin et al. (1999) find disturbed kinematics in $\sim$ 50\% of Virgo
galaxies, suggesting that tidal encounters are frequent in the
Virgo Cluster (see also Dale et al. 2001).
Although tidal interactions can clearly cause enhancements in the star formation
rate, there is not a one-to-one correspondence between
the strength of the tidal interactions and the star formation rate.
In the majority of tidal interactions, star formation enhancements are modest
(Kennicutt 1998).
This is consistent with the moderately enhanced star formation rates
observed in some of the Virgo galaxies in apparent pairs (Section~\ref{pair}).
Thus H$\alpha$ may not be a good tracer of all dynamical effects 
caused by collisions and tidal interactions.
Some of this is timescale:  the enhanced star-forming phase, if any,
may be short-lived, compared to the time that the galaxy is dynamically disturbed,
and experiencing significant radial mass transfer and dynamical heating of the disk.

\subsection{Gravitational and ICM-ISM Stripping?}

Many Virgo galaxies should be affected by
multiple types of interactions, since a large fraction
of spirals show evidence of stripping (this paper), 
and most Virgo spirals have either disturbed rotation curves (Rubin et al. 1999) 
or show some other evidence for tidal interactions.

While tidal interactions are clearly an important process driving cluster
galaxy evolution, it is doubtful that tidal interactions
alone could be responsible for most spiral-S0 transformations.
The combination of tidal interactions with ICM-ISM stripping
can drive the evolution of spirals toward a lenticular state.
Strong tidal interactions typically drive inner gas inward and outer
gas outward in tidal tails, depleting the gas at intermediate radii
(e.g., Barnes \& Hernquist 1991). 
Some of the gas driven toward the center forms stars,
yielding a galaxy with a higher central concentration of stars,
although not necessarily a bulge.
However, most tidal interactions will not clean out the outer disk of
star-forming gas, 
and this is probably required for a galaxy to be classified as lenticular.
ICM-ISM stripping seems to be required for the disk cleaning, although
tidal interactions can make a galaxy more susceptible to
ICM-ISM stripping by driving some of the outer disk gas 
outward, where it becomes easier to strip.
In this way a tidal interaction can facilitate the stripping of
the ISM by the ICM.
This combination of effects is likely for the truncated/compact 
H$\alpha$ class of galaxies including NGC~4064 and NGC~4424.
These spiral galaxies should become small bulge S0 galaxies 
within ~1 Gyr.

\subsection{Starvation}

Larson, Tinsley, \& Caldwell (1980) proposed a `starvation' model,
in which the normal gaseous inflow which fuels ongoing star formation in 
spirals
is cut off in clusters, as spirals fall into the cluster for the first time.
Abraham et al. (1996) and Balogh et al. (1999, 2000)
find that the evolution of the galaxies in several clusters
is consistent with 
a significant reduction in star formation rather than starbursts.
A reduction could be caused by either ICM-ISM stripping or starvation.
Starvation would be gradual whereas
ICM-ISM stripping would be relatively abrupt, but may not be complete
since a central gas disk can survive.
Based on the relatively small number of k+a galaxies,
with strong Balmer absorption lines but no emission lines,
Balogh et al. (1999, 2000) suggest
that the reduction in star formation is not abrupt and complete,
but is instead more gradual.
This would occur with starvation
(also called `strangulation' by Balogh \& Morris 2000),
but might also be consistent with abrupt partial ICM-ISM stripping.
Treu et al. (2003) find evidence for a possible difference in
the morphological mix of galaxies between the field and
the region outside the viral radius
(r$\sim$1-2 Mpc from the cluster center) of the 
in the z=0.4 cluster Cl0024+16,
and suggest that starvation and tidal effects may be 
significant processes in the outer cluster.

The simulations of Bekki et al. (2002) show that starvation
leads to gas and star formation distributions which closely
resemble anemic spirals, with lowered rates of star formation
throughout the disk. Gas infall, or its curtailment,
should not have a radial dependence strong enough to produce sharply truncated
gas disks. If the primary mechanism for gas loss were
starvation, we might therefore expect to see a large fraction of anemic 
spirals, particularly since starvation occurs on a much longer timescale
than disk gas stripping.

The fraction of purely anemic galaxies in the 2 environments is similar
(6\% Virgo, 4\% isolated).
8\% of the Virgo sample, but none of the isolated sample,
are classified as truncated/anemic.
The origin of the H$\alpha$ distributions in these 4 systems may be diverse
(Section~\ref{anemtrun}). Some appear tidally disturbed,
whereas others could be anemic disks that were truncated,
or spirals that were truncated a long time ago and have now faded into anemia.
Even if all the purely anemic and truncated/anemic galaxies
in Virgo are added together,
the total anemic fraction is 14\% in Virgo and 4\% isolated.
The fraction of anemic galaxies is somewhat higher in Virgo,
although low in both environments.
If starvation causes anemia, 
these results indicate that starvation is not the main cause
of reduced star formation in the Virgo cluster.

\subsection{Summary}

The spatial distributions of star formation in Virgo Cluster 
and isolated spirals provide evidence that both ICM-ISM stripping 
and tidal interactions have had a major influence 
on the evolution of Virgo Cluster spiral galaxies.

H$\alpha$ is a good tracer of 
active gas stripping and past partial gas stripping.
Gas stripping can significantly and irreversibly diminish
the future star formation potential of a galaxy,
and the truncated star forming disk in partially stripped galaxies
will likely remain apparent for more than 10$^9$ years.
Among the truncated/normal spiral galaxies, the most severe truncation
is 0.3$r_{24}$.
If a galaxy is stripped much more severely, it may be classified as an S0.
NGC~4710 is an edge-on Virgo S0 with abundant molecular gas
and star formation in the central 2 kpc = 0.2$r_{24}$ (Wrobel \& Kenney 1992),
and is a good candidate for an S0 created by severe but incomplete ICM-ISM stripping.
Fully stripped spirals become S0s, although
it is not possible to tell from its H$\alpha$ properties  
whether a galaxy became an S0 via complete stripping or some other process.
Since the most severely stripped spirals become S0's, which are
not in our sample, ICM-ISM stripping likely affects an even larger fraction 
of spirals than indicated by our study.

H$\alpha$ should also be a good tracer of starvation.
The future star formation potential of a galaxy is
greatly and probably irreversibly diminished by arrested gas infall, 
and the anemic star forming disk will likely remain apparent 
for more than 10$^9$ years.
Thus the large number of H$\alpha$-truncated galaxies compared to 
H$\alpha$-anemic galaxies should be a reflection of the 
greater importance of ICM-ISM stripping than starvation
for cluster galaxy evolution.

H$\alpha$ is certainly sensitive to some tidal effects,
but probably less sensitive to tidal effects than to gas stripping or starvation.
Many galaxies that experience tidal encounters experience
enhanced star formation, but only for brief intervals.
Other tidally interacting galaxies do not 
have greatly altered star formation rates (Kennicutt 1998),
yet they  probably experience radial gas inflow and disk heating 
from the tidal encounters.
For all but mergers and the strongest tidal encounters,
the long term star formation potential may not be greatly altered,
and in many cases, it will be hard to tell 10$^9$ years later
that the galaxy had a tidal interaction from the H$\alpha$ properties.
Thus in contrast to gas stripping or starvation,
we only see those galaxies which are presently experiencing tidal encounters
in H$\alpha$.
Overall, tidal effects 
are likely even more important for cluster galaxy evolution
than indicated by our H$\alpha$ study.

\section{Conclusions}

The results 
of our study of H$\alpha$ morphologies and cluster environmental effects
in Virgo spirals include the following:

1. About half of Virgo spirals have truncated H$\alpha$ disks,
which are relatively rare in isolated spirals (52\% Virgo vs. 12\% isolated).
A small fraction of Virgo spirals are anemic (6\% Virgo vs. 4\% isolated),
or truncated plus anemic (8\% Virgo vs. 0\% isolated),
or have enhanced  star formation (8\% Virgo vs. 0\% isolated).
37\% of the Virgo disks and 83\% of the isolated spirals 
are classified as normal.

2. The cluster locations vary for the  H$\alpha$ classes.
Normal galaxies are virtually absent from the cluster core.
Truncated galaxies are strongly concentrated in the core,
but some are also found in the cluster outskirts.
Anemic galaxies and those with enhanced NMSFRs are found throughout the
cluster.

3. The widespread spatial truncation of H$\alpha$ disks and greater
concentration of galaxies with truncated H$\alpha$ disks toward the 
cluster core provide evidence
that ICM-ISM stripping is primarily responsible for the overall
reduction of SFRs for Virgo spiral galaxies.

4. The fraction of anemic galaxies is modest in both environments,
suggesting that starvation is not a major factor in the
reduction of SFRs for Virgo spiral galaxies.

5. Several spirals have asymmetric H$\alpha$ enhancements at the outer edge of
truncated  H$\alpha$ disks, 
and at least two highly inclined spirals have extraplanar concentrations 
of HII regions, both suggestive of active ICM pressure.
These galaxies are located as far as 4.5$^{\circ}$ $\sim$1 Mpc from M87.
Their local star formation enhancements 
appear to cause only modest enhancements in the global star formation rates.

6. Three galaxies  with luminosities of 0.2-0.4L$^*$ and one with 2L$^*$
have global and inner NMSFRs 2-3 times higher than any isolated 
sample galaxy.
All are HI normal to HI rich, and are in apparent binary pairs and/or 
associated with large, extended clouds of HI, 
suggesting that tidal interactions and perhaps HI accretion
within groups of galaxies falling into the cluster for the first time,
are responsible for enhanced NMSFRs in an 8\% minority of Virgo spirals.

7. There is a clear correlation, although not a perfect one-to-one 
correspondence,
between the HI and H$\alpha$ radial distributions of large Virgo spirals.
The galaxies with severely truncated HI disks also have 
the most truncated H$\alpha$ disks.
The most anemic galaxies in H$\alpha$ are generally those with low HI surface
densities across the disk.
Galaxies with more peculiar HI morphologies also tend to have peculiar
H$\alpha$ morphologies.

8. At least two classes of galaxies with severely truncated star-forming
disks are identified. The most common type shows `simple truncation',
in which the inner disk star formation rates are
similar to those in low $C30$ isolated galaxies, but there is no star
formation beyond 0.3-0.4 $r_{24}$.  This
peculiar morphology has resulted in several cases in the assignment
of a mixed Hubble type such as Sc/Sa, and in other cases an early spiral
Hubble type. The morphology of these galaxies is most likely due to
ICM-ISM interactions. 
The other type of severely truncated galaxies shows
compact, non-axisymmetric, circumnuclear star formation.
These properties are more likely due to gravitational
interactions, including mergers.

9. These results support evidence that many of the lenticulars in z$<$0.1
rich clusters may indeed be stripped spirals, by demonstrating
an intermediate population of small bulge spirals which are
partially stripped, and misleadingly classified as Sa's or other early types.
Spirals which are stripped more severely and completely,
as is likely in clusters with denser ICMs and higher velocity
dispersions than in Virgo, would naturally be classified as lenticulars.
In saying this we do not wish to imply that ICM-ISM stripping
is the only mechanism driving lenticular formation in clusters.
Some of the galaxies in Virgo with truncated H$\alpha$ disks are
clearly much more peculiar than could be produced by ICM-ISM stripping,
and have likely experienced tidal encounters or mergers.
These peculiar systems are very HI-deficient and
exhibit no detectable outer disk star formation, and have
likely been stripped by the ICM in addition to experiencing
a tidal encounter.
In Virgo these are less common than the purely ISM-stripped spirals,
but the ratio could well be different in other clusters.

We thank the referee Alessandro Boselli for his detailed comments which helped
strengthen the paper.
The funding for the research on the Virgo cluster
and isolated spiral galaxies was provided by NSF grants AST-9322779 and
AST-0071251.
This research has made use of the NASA/IPAC Extragalactic Database (NED)
which is operated by the Jet Propulsion Laboratory, California Institute
of Technology, under contract with the National Aeronautics and Space
Administration.

\begin{deluxetable}{lcccrrrrrccl}
\tabletypesize{\scriptsize}
\tablecaption{H$\alpha$ Properties and Morphologies}
\tablewidth{0pt}
\tablehead{
\colhead{(1)}&
\colhead{(2)}&
\colhead{(3)} &
\colhead{(4)} &
\colhead{(5)} &
\colhead{(6)} &
\colhead{(7)} &
\colhead{(8)} &
\colhead{(9)} &
\colhead{(10)} &
\colhead{(11)} &
\colhead{(12)} \\
\colhead{Name}&
\colhead{Star }&
\colhead{RSA/BST}&
\colhead{RC3}&
\colhead{$M_B$}&
\colhead{v$_{he}$}&
\colhead{D$_{87}$}&
\colhead{HI }&
\colhead{CH$\alpha$}&
\colhead{EW}&
\colhead{EW }&
\colhead{Comments on}\\
\colhead{}&
\colhead{Formation}&
\colhead{}&
\colhead{}&
\colhead{}&
\colhead{(km s$_{-1}$}&
\colhead{($^{\circ}$)}&
\colhead{Def}&
\colhead{}&
\colhead{tot}&
\colhead{in}&
\colhead{H$\alpha$ or Stellar}\\
\colhead{}&
\colhead{Class}&
\colhead{}&
\colhead{}&
\colhead{}&
\colhead{}&
\colhead{}&
\colhead{}&
\colhead{}&
\colhead{(\AA)}&
\colhead{(\AA)}&
\colhead{Morphology}}
\startdata
NGC 4064& T/C&SBc(s):   & SB(s)a:pec &-18.72 &  913&8.8& 1.23 &0.97&8&19&\\
NGC 4178& N &SBc(s)II & SB(rs)dm     &-19.13 &  378&4.7&-0.09 &0.15&44&26&EEH\\
NGC 4189& N &SBc(sr)II.2& SAB(rs)cd?   &-18.49 & 2115&4.3& 0.27 &0.15&27&14&EEH\\
NGC 4192& N & SbII:   & SAB(s)ab     &-20.10 & -142&4.8& 0.18 &0.17&13&7&\\
NGC 4212& T/N & Sc(s)II-III& SAc:    &-19.16 & -81 &4.0& 0.16 &0.37&18&17&\\
NGC 4237& T/N & Sc(r)II.8& SAB(rs)bc &-18.65 &  867&4.4& 0.50 &0.55&10&11&\\
NGC 4254& N & Sc(s)I.3   & SA(s)c    &-20.59& 2407&3.6&-0.14 &0.32&36&25&\\
NGC 4293& T/A &Sa pec& (R)SB(s)0/a   &-19.82&  893&6.4& 1.45 &1.00&1&4& tidally disturbed\\
NGC 4294& N & SBc(s)II-III&SB(s)cd   &-18.40 &  355&2.5&-0.37 &0.42&47&61& P (NGC 4299), EEH\\
NGC 4298& T/N & Sc(s)III   & SA(rs)c &-18.94 & 1135&3.2& 0.08 &0.45&13&16&P (NGC 4302)\\
NGC 4299& T/E &Scd(s)III  & SAB(s)dm:&-18.16 &  232&2.4&-0.22 &0.46&84&117&P (NGC 4294)\\
NGC 4303& E &Sc(s)I.2   & SAB(rs)bc  &-20.85 & 1566&4.8&-0.03 &0.49&61&62&\\
NGC 4321& T/N & Sc(s)I     & SAB(s)bc&-20.91 & 1571&3.9& 0.21 &0.46&18&21&\\
NGC 4351& T/N(s)& Sc(s)II.3&SB(rs)ab:pec&-17.98&2310&1.7&0.37 &0.77&16&32&off-center H$\alpha$\\
NGC 4380& T/A &Sab(s)II-III& SA(rs)b:  &-18.66 &  967&2.7& 0.66 &0.77&4&5& EEH\\
NGC 4383& E &Amorph     & Sa? pec     &-18.34 & 1710&4.3&-0.53 &0.84&73&89&\\
NGC 4394& A&SBb(sr)I-II& (R)SB(r)b    &-19.26 &  922&5.9& 0.39 &0.25&5&3&\\
NGC 4405& T/N (s)& Sc(s)/S0& SA(rs)0/a:  &-18.03 & 1747&4.0& 0.85  &0.91&10&19&EEH\\
NGC 4411B& N &Sc(s)II    & SAB(s)cd     &-18.10 & 1270&3.6& 0.60 &0.35&20&22&\\
NGC 4413&  T/N &SBbc(rs)II-III& SB(rs)ab:&-18.05 &  102&1.1& 0.26 &0.58&16&25&\\
NGC 4419& T/A &Sa         & SB(s)a  &-18.89 & -261&2.8& 0.78 &0.64&2&3&\\
NGC 4424& T/C &Sa pec     & SB(s)a: &-18.70 &  439&3.1& 0.79 &1.00&10&24&merger\\
 IC 3392& T/N (s)&Sc/Sa   & SAb:  &-17.72 & 1687&2.7& 0.96 &0.82&7&17&\\
NGC 4450& T/A &Sab pec    & SA(s)Aab     &-20.09 & 1954&4.7& 0.99 &0.69&2&3&\\
NGC 4457& T/N (s)& RSb(rs)II&(R)SAB(s)0/a&-19.26&882&8.8&0.92 &0.66&8&12&one strong H$\alpha$ arm\\
NGC 4498&  N &SBc(s)II   & SAB(s)d      &-18.40 & 1507&4.5& 0.16 &0.25&31&24&\\
NGC 4501&  T/N &Sbc(s)II   & SA(rs)b     &-20.75 & 2281&2.0& 0.34 &0.44&15&15&\\
NGC 4519& N & SBc(rs)II.2& SB(rs)d       &-18.68 & 1220&3.8&-0.18 &0.14&52&20&P (NGC 4519A)\\
NGC 4522&  T/N (s)&Sc/Sb:& SB(s)cd:pec  &-18.29 & 2328&3.3& 0.69 &0.56&17&34& extraplanar HII\\
NGC 4527& N & Sb(s)II    & SAB(s)bc    &-19.70 & 1736&9.8&-0.19 &0.29&27&19& \\
NGC 4532& E &SmIII      & IBm        &-18.72 & 2012&6.0&-0.35 &0.82&105&115& P (DDO137)\\
NGC 4535& N & SBc(s)I.3  & SAB(s)c      &-20.51 & 1961&4.3& 0.16 &0.17&28&17&\\
NGC 4536& N & Sc(s)I     & SAB(rs)bc     &-20.01 & 1804&10.2&0.17 &0.46&23&28&\\
NGC 4548& A &SBb(rs)I-II& SB(rs)b        &-20.04 &  486&2.4& 0.76 &0.11&4&1&\\
NGC 4561& N & SBcIV      & SB(rs)dm      &-18.06:& 1407&7.0&-0.64 &0.30&43&33&\\
NGC 4567&  T/N &Sc(s)II-III& SA(rs)bc    &-18.94 & 2274&1.8& 0.64 &0.55&16&20& P (NGC 4568)\\
NGC 4568&  T/N &Sc(s)III   & SA(rs)bc    &-19.32 & 2255&1.8& 0.64 &0.62&23&32&P (NGC 4567)\\
NGC 4569& T/N (s)&Sab(s)I-II & SAB(rs)ab &-20.77 & -235&1.7& 0.92 &0.92&4&9& extraplanar HII\\
NGC 4571& N & Sc(s)II-III& SA(rs)d       &-19.21 &  342&2.4& 0.49 &0.27&13&10&\\
NGC 4579&  T/N &Sab(s)II   & SAB(rs)b    &-20.46 & 1519&1.8& 0.63 &0.61&8&9&bl T/A\\
NGC 4580&  T/N (s)&Sc/Sa   & SAB(rs)a pec&-18.53 & 1034&7.2& 1.31  &0.93&5&11&\\
NGC 4606& T/C&Sa pec     & SB(s)a:      &-18.33 & 1664&2.5& 1.16 &0.53&4&8&P? (NGC 4607)\\
NGC 4639& N & SBb(r)II   & SAB(rs)bc    &-18.83 & 1010&3.1& 0.18 &0.25&19&10&\\
NGC 4647& N & Sc(rs)III  & SAB(rs)c      &-18.99 & 1422&3.2& 0.43 &0.44&24&27&P (NGC 4649), EEH\\
NGC 4651& N & Sc(r)I-II  & SA(rs)c       &-19.66 &  805&5.1&-0.28 &0.56&22&21&minor merger\\
NGC 4654& N & SBc(rs)II  & SAB(rs)cd    &-19.88 & 1035&3.3& 0.05 &0.30&22&18&EEH\\
NGC 4689&  T/N &Sc(s)II.3  & SA(rs)bc   &-19.47 & 1616&3.7& 0.57 &0.67&10&17&\\
NGC 4694& T/N &Amorph     & SB0 pec  &-18.83 & 1175&4.5& 0.81 &0.95&3&4&bl T/A\\
NGC 4698& A& Sa         & SA(s)ab       &-19.49 & 1002&5.8&-0.14 &0.31&4&1&merger\\
NGC 4713& N & SBc(s)II-III& SAB(rs)d     &-18.81 &  653&8.5&-0.35 &0.28&59&45&\\
NGC 4772&  T/N &...        & SA(s)a  &-19.13 & 1040&11.7&0.07 &0.43&6&5&merger, bl T/A\\
NGC 4808& N & Sc(s)III   & SA(s)cd:     &-18.46:&  766&10.2&-0.68&0.24&43&27&\\
\enddata
\label{assign}
\tablecomments{
(1) Name of galaxy,
(2) Star Formation Class, where N=Normal, A=Anemic, E=Enhanced, 
T/A=Truncated/Anemic, T/C=Truncated/Compact, T/E=Truncated/Enhanced, 
T/N=Truncated/Normal, T/N(s)=Truncated/Normal (severe),
(3) Hubble types from BST or Sandage \& Tammann (1987) or
Sandage \& Bedke (1994),
(4) Hubble type from deVaucouleurs et al. (1991),
as provided by NASA/IPAC Extragalactic Database (NED),
(5) the total, face-on absolute blue magnitude ($M_B$), derived
from deVaucouleurs et al. (1991), assuming a distance of 16 Mpc,
(6) the heliocentric radial velocity,
(7) the projected angular distance in degrees of the galaxy from M87,
(8) the HI deficiency parameter, which was
calculated as described in PIII,
(9) the central H$\alpha$ concentration,
(10) the H$\alpha$ equivalent width, calculated as in PIII,  
over the whole disk,  
(11) the H$\alpha$ equivalent width, calculated as in PIII,  
over the inner 0.3$r_{24}$ disk,
(12) comments about galaxy characteristics, such as the presence
of edge-enhanced H$\alpha$ emission (EEH), borderline T/A (bl T/A), membership in
a pair (P(name of pair galaxy)), and/or evidence of a merger or tidal interaction. 
See text for references.}
\end{deluxetable}

\end{document}